\PassOptionsToPackage{unicode}{hyperref}
\PassOptionsToPackage{hyphens}{url}
\PassOptionsToPackage{dvipsnames,svgnames,x11names}{xcolor}
\documentclass[
]{article}
\usepackage{amsmath,amssymb}
\usepackage{iftex}
\ifPDFTeX
  \usepackage[T1]{fontenc}
  \usepackage[utf8]{inputenc}
  \usepackage{textcomp} % provide euro and other symbols
\else % if luatex or xetex
  \usepackage{unicode-math} % this also loads fontspec
  \defaultfontfeatures{Scale=MatchLowercase}
  \defaultfontfeatures[\rmfamily]{Ligatures=TeX,Scale=1}
\fi
\usepackage{lmodern}
\ifPDFTeX\else
\fi
\IfFileExists{upquote.sty}{\usepackage{upquote}}{}
\IfFileExists{microtype.sty}{% use microtype if available
  \usepackage[]{microtype}
  \UseMicrotypeSet[protrusion]{basicmath} % disable protrusion for tt fonts
}{}
\makeatletter
\@ifundefined{KOMAClassName}{% if non-KOMA class
  \IfFileExists{parskip.sty}{%
    \usepackage{parskip}
  }{% else
    \setlength{\parindent}{0pt}
    \setlength{\parskip}{6pt plus 2pt minus 1pt}}
}{% if KOMA class
  \KOMAoptions{parskip=half}}
\makeatother
\usepackage{xcolor}
\usepackage[margin=1in]{geometry}
\usepackage{color}
\usepackage{authblk}
\usepackage{fancyvrb}

\DefineVerbatimEnvironment{Highlighting}{Verbatim}{commandchars=\\\{\}}
\usepackage{framed}
\definecolor{shadecolor}{RGB}{248,248,248}
\newenvironment{Shaded}{\begin{snugshade}}{\end{snugshade}}

\newcommand{\AttributeTok}[1]{\textcolor[rgb]{0.13,0.29,0.53}{#1}}

\newcommand{\ConstantTok}[1]{\textcolor[rgb]{0.56,0.35,0.01}{#1}}

\newcommand{\DecValTok}[1]{\textcolor[rgb]{0.00,0.00,0.81}{#1}}

\newcommand{\FloatTok}[1]{\textcolor[rgb]{0.00,0.00,0.81}{#1}}
\newcommand{\FunctionTok}[1]{\textcolor[rgb]{0.13,0.29,0.53}{\textbf{#1}}}

\newcommand{\NormalTok}[1]{#1}

\newcommand{\OtherTok}[1]{\textcolor[rgb]{0.56,0.35,0.01}{#1}}

\newcommand{\SpecialCharTok}[1]{\textcolor[rgb]{0.81,0.36,0.00}{\textbf{#1}}}

\newcommand{\StringTok}[1]{\textcolor[rgb]{0.31,0.60,0.02}{#1}}

\usepackage{graphicx}
\makeatletter
\def\maxwidth{\ifdim\Gin@nat@width>\linewidth\linewidth\else\Gin@nat@width\fi}
\def\maxheight{\ifdim\Gin@nat@height>\textheight\textheight\else\Gin@nat@height\fi}
\makeatother
\setkeys{Gin}{width=\maxwidth,height=\maxheight,keepaspectratio}
\makeatletter
\def\fps@figure{htbp}
\makeatother
\providecommand{\tightlist}{%
  \setlength{\itemsep}{0pt}\setlength{\parskip}{0pt}}
\newcommand{\norm}[1]{\left\lVert #1 \right\rVert}
\newcommand{\snorm}[1]{\lVert #1 \rVert}
\newcommand{\R}{\mathbb{R}}
\newcommand{\vbeta}{\boldsymbol\beta}
\newcommand{\vx}{\mathbf{x}}
\newcommand{\mX}{\mathbf{X}}
\newcommand{\vY}{\mathbf{y}}
\newcommand{\vy}{\vY}
\newcommand{\df}{\mathrm{df}}
\newcommand{\vR}{\mathbf{r}}
\newcommand{\Hessian}{\mathbf{H}}
\usepackage{amsmath, amsfonts, amsthm, amssymb}
\usepackage{algorithm, algorithmic}
\usepackage{nicefrac}
\usepackage{booktabs}
\usepackage[sort&compress,round]{natbib}
\usepackage[margin=1in]{geometry}
\usepackage{xcolor}

\def\T{\mathsf{T}}

\usepackage[pagebackref=true]{hyperref}
\hypersetup{colorlinks=true,citecolor=blue,linkcolor=blue,urlcolor=blue}

\newtheorem{proposition}{Proposition}

\newcommand\code{\bgroup\@makeother\_\@makeother\~\@makeother\$\@codex}
\def\@codex#1{{\normalfont\ttfamily\hyphenchar\font=-1 #1}\egroup}
\let\code=\texttt
\let\proglang=\textsf
\newcommand{\pkg}[1]{{\fontseries{m}\fontseries{b}\selectfont #1}}
\newcommand{\email}[1]{\href{mailto:#1}{\normalfont\texttt{#1}}}
\ifLuaTeX
  \usepackage{selnolig}  % disable illegal ligatures
\fi
\IfFileExists{bookmark.sty}{\usepackage{bookmark}}{\usepackage{hyperref}}
\IfFileExists{xurl.sty}{\usepackage{xurl}}{} % add URL line breaks if available
\hypersetup{
  pdftitle={: An  Package for Estimating Sparse Group Lasso},
  colorlinks=true,
  linkcolor={DarkBlue},
  filecolor={Maroon},
  citecolor={DarkBlue},
  urlcolor={DarkBlue},
  pdfcreator={LaTeX via pandoc}}

\title{\pkg{sparsegl}: An \proglang{R} Package for Estimating Sparse
Group Lasso}

\author[a]{Xiaoxuan Liang}
\author[b]{Aaron Cohen}
\author[c]{Anibal Sol\'on Heinsfeld}
\author[d,e]{Franco Pestilli}
\author[a,*]{Daniel J. McDonald}

\affil[a]{Department of Statistics, The University of British Columbia, Vancouver, BC Canada}
\affil[b]{Department of Statistics, Indiana University, Bloomington, IN USA}
\affil[c]{Department of Computer Science, The University of Texas, Austin TX USA}
\affil[d]{Department of Psychology, The University of Texas, Austin TX, USA}
\affil[e]{Center for Perceptual Systems, The University of Texas, Austin TX, USA}

\date{}

\begin{document}
\maketitle
\footnotetext[1]{To whom correspondence should be
  addressed. E-mail: \email{daniel@stat.ubc.ca}} 
\begin{abstract}
The sparse group lasso is a high-dimensional regression technique that
is useful for problems whose predictors have a naturally grouped
structure and where sparsity is encouraged at both the group and
individual predictor level. In this paper we discuss a new \proglang{R}
package for computing such regularized models. The intention is to
provide highly optimized solution routines enabling analysis of very
large datasets, especially in the context of sparse design matrices.
\end{abstract}

\hypertarget{introduction}{%
\section{Introduction}\label{introduction}}

\shortcites{tibshirani2012strong,vaiter2012degrees,R-glmnet,pestilli2014,aminmansour2019,Schiavi2020,vanessen2012,R-ggplot2,R-rticles}

\nocite{R-dplyr,R-magrittr}

Regularized linear models are now ubiquitous tools for prediction and,
increasingly, inference. When solving such high-dimensional learning
problems, adding regularization helps to reduce the chances of
overfitting and improve the model performance on unseen data. Sparsity
inducing \(\ell_1\)-type penalties such as the lasso
\citep{tibshirani1996regression} or the Dantzig selector
\citep{CandesTao2007} perform both variable selection and shrinkage,
resulting in near-optimal statistical properties. The group lasso
\citep{yuan2006model} modifies the regularizer, replacing the \(\ell_1\)
penalty with a groupwise sum of \(\ell_2\) norms. When covariates have
natural groupings, such as with genomics data or one-hot encoded
factors, this group penalty is preferable, because the resulting
estimate will include or exclude entire groups of covariates. To
simultaneously attain sparsity at both group and individual feature
levels, \citet{simon2013sparse} proposed the sparse group-lasso, a
convex combination of the \(\ell_1\) lasso penalty and the group lasso
penalty.

While a number of packages exist for solving the sparse group lasso, our
\proglang{R} \citep{R-base} implementation in \pkg{sparsegl}
\citep{R-sparsegl} is designed to be fast, especially in the case of
large, sparse covariate matrices. This package focuses on finding the
optimal solutions to sparse group-lasso penalized learning problems at a
sequence of regularization parameters, implements risk estimators in an
effort to avoid cross validation if necessary, leverages a fast,
compiled \proglang{Fortran} implementation, avoids extraneous data
copies, and undertakes a number of additional computational efficiency
improvements. In \proglang{R}, there are already excellent
implementations of sparse group lasso and group lasso, namely \pkg{SGL}
\citep{R-SGL}, \pkg{gglasso} \citep{R-gglasso,yang2015fast}, and
\pkg{biglasso} \citep{R-biglasso,zeng2020biglasso}. Of these, only
\pkg{SGL} employs the additional \(\ell_1\) sparsity-inducing penalty.
However, it has a number of drawbacks that result in much slower
performance, even on small data. One major reason is the omission of
so-called ``strong rules'' \citep{tibshirani2012strong} that help
coordinate descent algorithms to avoid many of the groups which will
turn out to have zero coefficient estimates. The \pkg{gglasso} and
\pkg{biglasso} packages are both computationally fast. The former
incorporates the strong rule, and the latter involves a hybrid
safe-strong rule along with scalable storage and a parallel
implementation in \proglang{C++} and \proglang{R} that allows for data
that exceeds the size of installed random access memory. Unfortunately,
neither allows within-group sparsity (i.e., they perform group lasso,
not sparse group lasso). Thus, the estimated coefficients produced by
these packages will have some active groups and some inactive groups,
but within an active group, generally all the coefficients will be
nonzero.

In \proglang{python} \citep{python}, \pkg{asgl}
\citep{civieta2020adaptive} implements adaptive sparse group-lasso,
which flexibly adjusts the weights in the penalization on the groups of
features. Additionally it incorporates quantile loss. As with the other
packages mentioned above, it can also solve the special cases (lasso and
group lasso). However, for all optimization problems, it directly uses
\pkg{CVXPY} \citep{cvxpy-jmlr,cvxpy-jcd}, a general purpose optimizer,
without strong rules or other tricks to relate solutions to each other
across values of the tuning parameter.\footnote{A similar implementation
  could be achieved in \proglang{R} using \pkg{CVXR}
  \citep{CVXR,R-CVXR}.}

Our contribution, then, is to provide a package that performs sparse
group lasso and is faster than existing implementations. In particular,
\pkg{sparsegl} has the following benefits:

\begin{itemize}
\item Performs Gaussian and logistic regression using fast, compiled code;
\item Allows arbitrary generalized linear models using \proglang{R}'s \code{family} object, though with slightly less efficiency;
\item Allows for interval constraints and differential weights on the coefficients;
\item Accommodates a sparse design matrix and returns the coefficient estimates in a sparse matrix;
\item Uses strong rules (and active set iteration) for fast computation along a sequence of tuning parameters;
\item Uses \pkg{dotCall64} to interface with low-level \proglang{Fortran} functions and avoid unnecessary copying as well as allow for 64-bit integers \citep[see][]{gerber2017dotcall,gerber2018dotcall};
\item Provides information criteria as risk estimators (AIC/BIC/GCV) in addition to cross validation.
\end{itemize}

\begin{figure}[t!]

{\centering \includegraphics{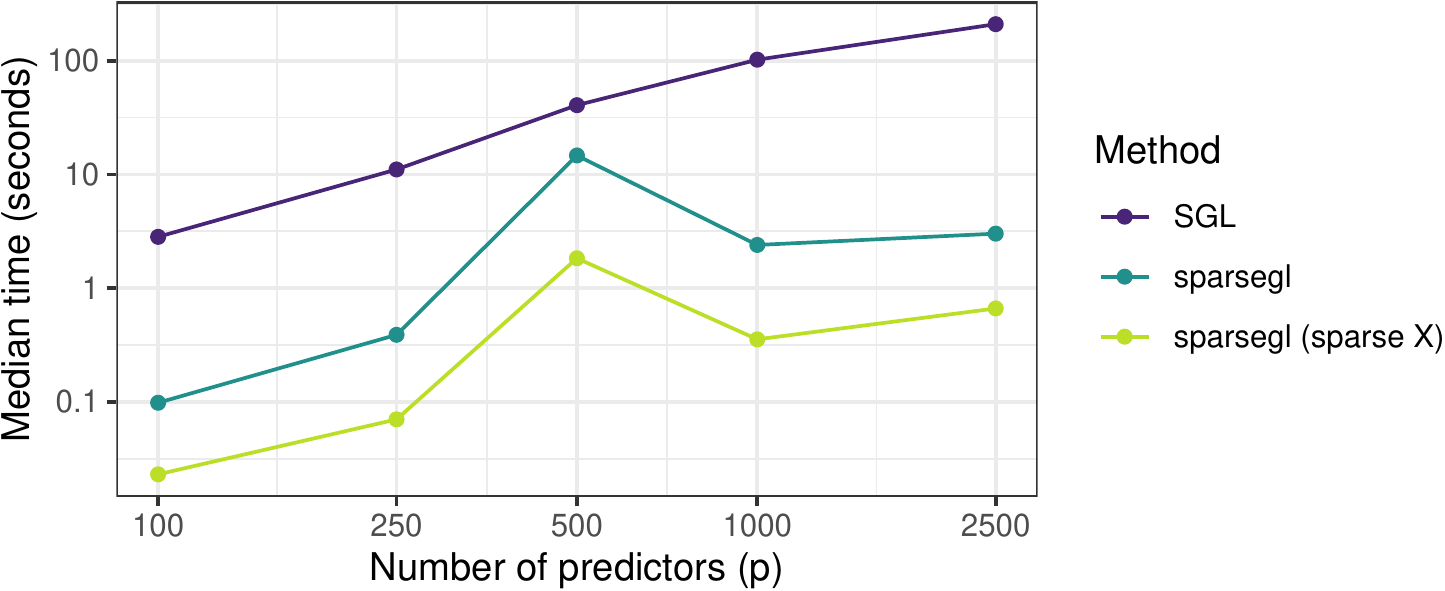} 

}

\caption{This figure shows the time required to compute sparse group lasso solutions across a number of different problem sizes. In all cases, we use $n=500$ observations and 100 values of the tuning parameter $\lambda$. The median is taken across 5 replications for each method and problem size. Note that both axes are on the log scale.}\label{fig:timing-comparison}
\end{figure}

A comparison of features of this and related \proglang{R} packages is
shown in \autoref{tab:comparison}. \autoref{fig:timing-comparison}
compares the speed of the dense and sparse implementations in
\pkg{sparsegl} with \pkg{SGL} across a number of different problem
sizes, finding speedups of 1.5 to 2.5 orders of magnitude.

In \autoref{methodology-estimation-and-prediction}, we describe the
algorithmic implementation in detail, paying particular attention to the
strong rule. In \autoref{example-usage}, we show how to use the package,
running through an example with simulated data. \autoref{applications}
demonstrates many of the unique features of \pkg{sparsegl} in two
applications. We summarize our contributions in \autoref{discussion}.

\begin{table}
\centering
\begin{tabular}{lcccccc}
\toprule
& Regression \& & Within group & Sparse & Strong & Avoids & Interval \\
& classification & sparsity & matrices & rules & copies & constraints \\
\midrule
\pkg{sparsegl} & \checkmark & \checkmark & \checkmark & \checkmark & \checkmark & \checkmark \\
\pkg{gglasso} & \checkmark & & &\checkmark \\
\pkg{SGL} & \checkmark & \checkmark \\
\bottomrule
\end{tabular}
\caption{This table summarizes the features available in \pkg{sparsegl} and related \proglang{R} packages.\label{tab:comparison}}
\end{table}

\hypertarget{methodology-estimation-and-prediction}{%
\section{Methodology, estimation and
prediction}\label{methodology-estimation-and-prediction}}

Given a sample of \(n\) observations of a univariate response \(y_i\)
and a corresponding vector of features \(\mathbf{x}_i \in \R^p\), the
standard linear regression setup has 
\begin{equation*}
\label{eq:linmod}
y_i = \mathbf{x}_i^\T \vbeta + \sigma\epsilon_i,\ \ i=1,\ldots,n,
\end{equation*} 
where \(\epsilon_i\) is independent standard Gaussian
noise and \(\sigma > 0\). While ordinary least squares estimates the
coefficient vector \(\vbeta\) by solving
\(\min_{\vbeta} \frac{1}{2n}\sum_{i=1}^n (y_i - \vx_i^\T\vbeta)^2\),
this method tends to behave poorly if \(p \gg n\). In what follows, we
will write \(\vy = (y_1,\ldots,y_n)\) and let \(\mX\) be the rowwise
concatenation of \(\vx^\T_1,\ldots,\vx_n^\T\).

The lasso adds an \(\ell_1\) penalty to the optimization problem:
\begin{equation}
\label{eq:lasso}
\min_{\vbeta} \left\{\frac{1}{2n} \sum_{i=1}^n (y_i - \vx_i^\T\vbeta)^2 + \lambda\sum_{j=1}^p |\beta_j|\right\} = \min_{\vbeta}\left\{\frac{1}{2n}\norm{\vY-\mX\vbeta}_2^2 + \lambda \norm{\vbeta}_1\right\},
\end{equation} where \(\norm{\cdot}_2\) is the Euclidean (\(\ell_2\))
norm and \(\norm{\cdot}_1\) is the \(\ell_1\) norm. The benefit of this
penalty is that it tends to allow only a subset of coefficient estimates
to be nonzero, hence performing variable selection. Here, \(\lambda\) is
a hyperparameter that trades fidelity to the data---small \(\lambda\)
emphasizes minimization of the squared-error---with desirable
regularization that selects a subset of variables and improves
prediction accuracy.

A variant of this, the group lasso \citep{yuan2006model} is appropriate
when there is a natural grouping structure for the features. That is, we
assume that both the design matrix \(\mX\) and the corresponding vector
of coefficients can be partitioned into interpretable non-overlapping
groups, and, by analogy with lasso regression, only a few of the groups
are active, i.e., have nonzero coefficients. The group lasso thus
performs regularization that has the effect of discarding groups of
predictors rather than the predictors themselves: \begin{equation}
\label{eq:group-lasso}
\min_{\vbeta}\left\{\frac{1}{2n}\big\lVert \vY-\sum_{g=1}^G \mX^{(g)}\vbeta^{(g)}\big\rVert_2^2 + \lambda\sum_{g=1}^G\sqrt{w_g}\snorm{\vbeta^{(g)}}_2\right\}.
\end{equation} Grouping may occur naturally---say with the inclusion of
many categorical predictors, groups of genes, or brain regions---or may
be a design choice using additive models and basis expansions. Note that
in Equation~\ref{eq:group-lasso}, the grouping structure is explicitly
stated: the vector of coefficients, \(\vbeta\), is thought of as a
concatenation of the coefficient subvectors of the various groups
\(\vbeta^{(g)}\), and similarly the data matrix \(\mX\) is the
concatenation of submatrices, each submatrix \(\mX^{(g)}\) being
composed of the columns that correspond to that particular group. Thus
the first part of the equation,
\(\vY-\sum_{g=1}^G\mX^{(g)}\vbeta^{(g)}\), is identical to the more
simply-written equation \(\vY-\mX\vbeta\), but the notation serves to
emphasise the partitioning.

However, the penalty, \(\sum_{g=1}^G\sqrt{w_g}\snorm{\vbeta^{(g)}}_2\),
is different from the corresponding part in Equation~\ref{eq:lasso},
using instead the sum of the (non-squared) \(\ell_2\)-norms of the
coefficient vectors of the various groups. It is the
non-differentiability of this expression at \(\mathbf{0}\in\R^{|g|}\)
(with \(|g|\) meaning the size of group \(g\)) that accounts for the
group-discarding property of the solution, similar to the way that the
non-differentiability the absolute value at \(0\) is responsible for
discarding individual predictors in the lasso.

As with Equation~\ref{eq:lasso}, there is only a single tuning parameter
\(\lambda\), whose value determines the strength of regularization.
Within the second summation are the relative weights of the groups,
\(w_g\). These are often taken to be the size of the corresponding
group. For simplicity, this notation is suppressed below where the
meaning is clear.

Finally, in a group-structured problem as above, it may be desirable to
enforce sparsity not only among the groups but also within the groups.
The sparse group lasso \citep{simon2013sparse} does this by combining
the penalties in Equations~\ref{eq:lasso} and~\ref{eq:group-lasso}:
\begin{equation}
  \label{eq:sparsegl}
\min_{\vbeta}\left\{\frac{1}{2n}\snorm{\vY-\mX\vbeta}_2^2 + (1-\alpha)\lambda\sum_{g=1}^G \snorm{\vbeta^{(g)}}_2+\alpha\lambda\sum_{g=1}^G\snorm{\vbeta^{(g)}}_1\right\}.
\end{equation} There is now a second tuning parameter \(\alpha\), which
controls the relative emphasis of intra- versus inter-group sparsity in
the coefficient estimates. In Equation~\ref{eq:sparsegl}, we have chosen
to write the group dependence explicitly in the \(\ell_1\) component of
the penalty, but note that
\(\sum_{g=1}^G\snorm{\vbeta^{(g)}}_1 = \snorm{\vbeta}_1\). Similar to
the weights \(w_g\) in the group component, \pkg{sparsegl} also allows
individual predictor weights in the \(\ell_1\) component,
\(\sum_{j=1}^p \omega_j |\beta_j|\), but we suppress this generality for
clarity, setting \(\omega_j=1\) for all \(j\) below.

\hypertarget{the-group-wise-majorization-minimization-algorithm}{%
\subsection{The group-wise majorization-minimization
algorithm}\label{the-group-wise-majorization-minimization-algorithm}}

There is no closed-form solution to the optimization problem in
Equation~\ref{eq:sparsegl}, so we require a numerical procedure. Because
the problem is convex, a variety of methods may be used. The general
framework for our algorithm is the same as the majorized block-wise
coordinate descent algorithm developed in
\citep{yang2015fast, simon2013sparse}. What this means is that, for a
fixed value of \(\lambda\), we loop over the groups and update only
those coefficients while holding all other groups constant. In
particular, instead of using the exact Hessian to determine the step
size and direction in every update step, we update according to a
simpler expression that majorizes the objective.

For the rest of this section, we describe this majorization algorithm,
focusing on a particular group \(g\) and holding the coefficients for
all other groups fixed. We note here that, because the loss function in
Equation~\ref{eq:sparsegl} is differentiable and the penalty terms are
convex and separable (i.e., they can be decomposed into a sum of
functions each only involving a single group), this block coordinate
descent algorithm is guaranteed to converge to a global optimum
\citep{tseng2001convergence}.

To begin with, we introduce some notation. Let \[
\vR_{(-g)} = \vY - \sum_{k \neq g} \mX^{(k)} \vbeta^{(k)}
\] be the partial residual without group \(g\) where all the group fits
besides that of group \(g\) are subtracted from \(\vY\). With all other
groups held fixed, we aim to solve: \begin{equation}
    \label{eq:sparsegroupk}
\min_{\vbeta^{(g)}} \frac{1}{2n} \snorm{\vR_{(-g)}-\mX^{(g)}\vbeta^{(g)}}_2^2 + (1-\alpha)\lambda \snorm{\vbeta^{(g)}}_2 + \alpha \lambda \snorm{\vbeta^{(g)}}_1. 
\end{equation} In what follows, we will suppress the \((g)\) notation,
with the understanding that we are really referring to only the
\(g^\textrm{th}\) group of the coefficient vector and the partial
residual \(\vR_{(-g)}\). We will also define, the unpenalized loss
function \[
\ell (\vbeta) = \frac{1}{2n}\snorm{\vR - \mX\vbeta}_2^2, 
\] so that our objective function for the \(g^\textrm{th}\) group
becomes
\(\ell (\vbeta) + (1-\alpha)\lambda\norm{\vbeta}_2 + \alpha \lambda \norm{\vbeta}_1\),
and we are interested in finding an optimal value, \(\hat{\vbeta}\).
This enables the procedure to generalize easily to logistic loss or, in
principle, other exponential families.

Any global minimum must satisfy a subgradient equation, similar to a
first-derivative test for an optimum, except that \(\norm{\cdot}_2\) and
\(\norm{\cdot}_1\) are non-differentiable at \(\mathbf{0}\). For
Equation~\ref{eq:sparsegroupk} above, taking the subdifferential and
setting equal to zero gives us the following first-order condition:
\begin{equation}
  \label{eq:subgrad}
\nabla \ell(\vbeta) = (1-\alpha)\lambda \mathbf{u} + \alpha \lambda \mathbf{v},
\end{equation} where \(\mathbf{u}\) is the subgradient of
\(\norm{\vbeta}_2\) and \(\mathbf{v}\) is the subgradient of
\(\norm{\vbeta}_1\). The first is defined to be
\(\vbeta / \norm{\vbeta}_2\) if \(\vbeta\) is a nonzero vector, and is
any vector in the set \(\{\mathbf{u} : \norm{\mathbf{u}}_2 \leq 1 \}\)
otherwise; the second, \(\mathbf{v}\), is defined coordinate-wise as
\(v_j = \text{sign}(\beta_j)\) if \(v_j \neq 0\), and is any value
\(v_j \in \{v_j : |v_j| \leq 1 \}\) otherwise.

For the Gaussian case, the unpenalized loss \(\ell (\vbeta)\) is a
quadratic function in \(\vbeta\), so it is equal to its second order
Taylor expansion about any point \(\vbeta_0\) in the parameter space. We
thus start with the following equality for any given \(\vbeta_0\)
(recalling that \(\vbeta_0\) here is only for group \(g\)):\\
\begin{equation*}
\forall \vbeta,\ \vbeta_0,\quad \ell(\vbeta) = \ell(\vbeta_0)+(\vbeta - \vbeta_0)^\T\nabla \ell(\vbeta_0)+\frac{1}{2}(\vbeta - \vbeta_0)^\T \Hessian (\vbeta - \vbeta_0),
\end{equation*} where the gradient \(\nabla \ell\) is the first total
derivative of \(\ell\) (evaluated at \(\vbeta_0\)) and \(\Hessian\), the
Hessian, is the second total derivative. For
\(\ell(\vbeta) = \frac{1}{2n}\snorm{\vR - \mX\vbeta}_2^2\), a short
computation shows that the Hessian is
\(\Hessian = \frac{1}{n}\mX^\T \mX\).

For the large-scale problems motivating this work, the matrix \(\mX\) is
large, so computing \(\mX^\T \mX\), storing it in memory, or inverting
it, is computationally prohibitive. Instead, we replace this matrix with
a simpler one, \(t^{-1}\mathbf{I}\), a diagonal matrix with the value of
\(t\) selected to be such that this dominates the Hessian (in the sense
that \(t^{-1}\mathbf{I} - \Hessian\) is positive definite). For our
algorithm we choose the largest eigenvalue of the Hessian and use that
for \(t^{-1}\). Note that this eigenvalue must be computed for each
group \(g \in G\), but this computation is relatively simple using the
power method or other techniques as implemented with \pkg{RSpectra}
\citep{R-RSpectra}. This upper bound leads to the following inequality:
\begin{equation}
\forall \vbeta,\ \vbeta_0,\quad \ell(\vbeta) \leq \ell(\vbeta_0)+(\vbeta - \vbeta_0)^\T\nabla \ell(\vbeta_0)+\frac{1}{2t}(\vbeta - \vbeta_0)^\T  (\vbeta - \vbeta_0).
\label{eq:dominate}
\end{equation}

Replacing the loss function in the original minimization problem in
Equation~\ref{eq:sparsegroupk} with the right-hand side of
Equation~\ref{eq:dominate} leads to a majorized version of the original
problem \begin{equation}
\label{eq:Meq}
\ell(\vbeta_0)+(\vbeta - \vbeta_0)^\T\nabla \ell(\vbeta_0)+\frac{1}{2t}\norm{\vbeta_0-\vbeta}_2^2+ (1-\alpha)\lambda\norm{\vbeta}_2+\alpha\lambda\norm{\vbeta}_1,
\end{equation} which no longer involves operations with the Hessian
matrix.

As before, the optimal value for Equation~\ref{eq:Meq} is determined by
its subgradient equation, similar to that of Equation~\ref{eq:subgrad}:
\begin{equation*}
\frac{1}{t} \Big(\vbeta - \big(\vbeta_0 - t\nabla \ell(\vbeta_0)\big)\Big) +(1-\alpha)\lambda \mathbf{u} + \alpha \lambda \mathbf{v} = \mathbf{0},
\end{equation*} with \(\mathbf{u}\) and \(\mathbf{v}\) as defined above.
Solving this for \(\vbeta\) in terms of \(\vbeta_0\) results in the
following expression: \begin{equation}
\hat{\vbeta} = U(\vbeta_0) =
\left(1-\frac{t(1-\alpha)\lambda}{\norm{S(\vbeta_0-t\nabla \ell(\vbeta_0),\ t\alpha\lambda)}_2}\right)_+ S\big(\vbeta_0-t\nabla \ell(\vbeta_0),\ t\alpha\lambda\big),
\label{eq:updateStep}
\end{equation} where \((z)_+ = \max\{z,\ 0\}\) and \(S\) is the
coordinate-wise soft threshold operator, on a vector
\(\boldsymbol\gamma\) and scalar \(b\), \begin{equation*}
\label{softthresh}
(S(\boldsymbol\gamma,b))_j = \text{sign}(\gamma_j)(|\gamma_j| - b)_+,
\end{equation*} i.e., for each coordinate in the vector, it shrinks that
coordinate in magnitude by the amount \(b\), and sets it to zero if the
magnitude of that coordinate was smaller than \(b\) to begin with. It is
this soft-thresholding operation that encourages within-group sparsity.

An examination of Equation~\ref{eq:updateStep} shows that it is possible
for the entire group to be set to zero (made inactive) due to the (hard)
threshold operator \((\cdot)_+\) in the first part of the expression. It
is also possible for individual components of \(\vbeta^{(b)}\) to be
zeroed out by the coordinate-wise (soft) threshold operator \(S\).
Therefore, performing this update step tends to enforce coefficient
sparsity at both the group- and individual-level.

\begin{algorithm}[tb!]
  \caption{Sparse group lasso solution for fixed $\lambda$, regression version}
  \label{alg:sparsegl}
  \begin{algorithmic}[1]
    \STATE {\bfseries Input:} $\lambda \geq 0$, $\alpha \in [0,\ 1]$, set of groups $\mathcal{G}$, initial coefficients $\vbeta$, $\mathbf{r} = \vy-\mX\vbeta$
    \WHILE{Not converged}
    \FOR{$g \in \mathcal{G}$}
    \STATE Update $\vbeta^{(g)} = \left(1-\frac{t(1-\alpha)\lambda}{\snorm{S\big(\vbeta^{(g)}-t\nabla \ell(\vbeta^{(g)}),\ t\alpha\lambda\big)}_2}\right)_+ S\big(\vbeta^{(g)}-t\nabla \ell(\vbeta^{(g)}),\ t\alpha\lambda\big)$.
    \STATE Update $\mathbf{r} = \mathbf{r} - \mX^{(g)}\vbeta^{(g)}$.
    \ENDFOR
    \ENDWHILE 
    \RETURN{$\vbeta$}
  \end{algorithmic}
\end{algorithm}

Above, we have focused on the Gaussian linear model with
\(\ell(\vbeta) = \frac{1}{2n}\snorm{\vR - \mX\vbeta}_2^2\),
\(\nabla \ell(\vbeta) = -\frac{1}{n}\mX^\T (\vR - \mX\vbeta)\), and
\(\mathbf{H}\preceq t^{-1}\mathbf{I}\). In the case of logistic
regression, we use exactly the same procedure but with
\(\ell(\vbeta) = \frac{1}{n}\sum_{i}\log(1 + \exp\{-r_i\vx_i^\T\vbeta\})\),
\(\nabla \ell(\vbeta) = -\frac{1}{n}\sum_{i}y_i\vx_i^\top (1 + \exp\{-r_i\vx_i^\T\vbeta\})^{-1}\),
and \(\mathbf{H}(\vbeta) \preceq 4t^{-1}\mathbf{I}\). This procedure is
explicitly stated in \autoref{alg:sparsegl}. For other exponential
families (for example, Poisson, Gamma, or Probit regression), we provide
functionality to pass an \proglang{R} \texttt{family} object. These will
generally be much slower than the built-in families described above
because they require iteratively reweighted least squares as an outer
loop combined with inner majorization-minimization iterations as
described here.

While the procedure described so far solves Equation~\ref{eq:sparsegl}
for fixed choices of \(\lambda\) and \(\alpha\), the data analyst does
not typically know these ahead of time. Rather, we would like to solve
the problem for a collection of values of \(\lambda\) (and perhaps
\(\alpha\) as well). It turns out that the structure of this
optimization problem allows for some heuristics that can perform this
sequential optimization with a minimum of additional computational
resources, in some cases, solving Equation~\ref{eq:sparsegl} faster for
a sequence of values \(\lambda_m \in \{\lambda_1,\ldots,\lambda_M\}\)
than for a single choice \citep{tibshirani2012strong}. We describe our
implementation of this procedure next.

\hypertarget{sequential-strong-rule-kkt-conditions-and-active-set-iteration}{%
\subsection{Sequential strong rule, KKT conditions, and active set
iteration}\label{sequential-strong-rule-kkt-conditions-and-active-set-iteration}}

For any fixed value of \(\lambda\), many groups of coefficient estimates
will end up being equal to zero. If, somehow, we knew \emph{which}
groups, we could completely avoid visiting them in the blockwise
coordinate descent updates, and therefore avoid calculating
Equation~\ref{eq:updateStep} for those groups. This would significantly
speed up computations.

Re-examining Equation~\ref{eq:subgrad}, we can see that the first order
condition implies that, for each group \(g\), any solution must satisfy
\begin{equation}
\label{eq:subgrad-again}
\snorm{S(\nabla \ell(\vbeta_g),\ \lambda\alpha)}_2 \leq (1-\alpha)\lambda.
\end{equation} This is because, as \(\mathbf{u}\) is the subgradient of
\(\snorm{\vbeta_g}_2\), \(\snorm{\mathbf{u}}_2\leq 1\). Furthermore, if
\(\snorm{\mathbf{u}}_2 < 1\), then \(\hat{\vbeta}_j = \mathbf{0}\). In
the previous section, we used the sufficiency of the Karush-Kuhn-Tucker
(KKT) stationarity condition to derive a solution, while
Equation~\ref{eq:subgrad-again} is the necessary version. So given a
potential solution, it is easy to check its validity. Unfortunately,
this is not constructive.

The sequential strong rule \citep{tibshirani2012strong} begins from
Equation~\ref{eq:subgrad-again} and makes use of the fact that we are
solving for a sequence of parameters
\(\{\lambda_1 > \lambda_2 > \dots > \lambda_M\}\) rather than a single
value. At each \(\lambda_m\), we rely on the fact that we have already
solved the problem at \(\lambda_{m-1}\) and use this information to
quickly discard many groups of predictors. Without loss of generality,
for the rest of this section, assume that the problem has been solved
for \(\lambda_{m-1}\).

Define \(c_g(\lambda) = S(\nabla \ell(\vbeta_g),\ \lambda\alpha)\). Now,
we make the assumption that \(c_g(\lambda)\) is
\((1-\alpha)\)-Lipschitz, i.e., that \begin{equation*}
\label{eq:lipschitz}
\forall \lambda,\ \lambda^{\prime}\ \  \ \snorm{c_g(\lambda)-c_g(\lambda^{\prime})}_2 \leq (1-\alpha)|\lambda - \lambda'|.
\end{equation*} This Lipschitz assumption appears unintuitive, and in
fact, is not always true, but it turns out to be useful.

By Equation~\ref{eq:subgrad-again}, if we knew that
\(\snorm{c_g(\lambda_m)}_2 < (1-\alpha)\lambda_m\) then we could safely
ignore it. But we already have the solution at \(\lambda_{m-1}\). By the
triangle inequality (first) and the Lipschitz assumption (second), \[
\snorm{c_g(\lambda_m)}_2 \leq \snorm{c_g(\lambda_m) - c_g(\lambda_{m-1})}_2 +
\snorm{c_g(\lambda_{m-1})}_2 \leq (1-\alpha)(\lambda_{m-1} - \lambda_m) + \snorm{c_g(\lambda_{m-1})}_2.
\] We want to be able to assert that
\((1-\alpha)(\lambda_{m-1} - \lambda_m) + \snorm{c_g(\lambda_{m-1})}_2 \leq (1-\alpha)\lambda_m\),
allowing us to ignore group \(g\), and this assertion holds precisely
when \begin{equation*}
\label{eq:strong}
\snorm{c_g(\lambda_{m-1})}_2 \leq (1-\alpha)(2\lambda_m - \lambda_{m-1}).
\end{equation*}

Applying this logic to Equation~\ref{eq:Meq} gives the sequential strong
rule for the sparse group lasso: \begin{equation}
\label{eq:sgl-strong}
\snorm{S\big(\nabla \ell(\vbeta_g),\ t\alpha\lambda_{m-1}\big)}_2 \leq t(1-\alpha)(2\lambda_m - \lambda_{m-1}).
\end{equation} For more details in related settings, see
\citet{tibshirani2012strong}. If Equation~\ref{eq:sgl-strong} holds,
then we ignore group \(g\) when solving the problem at \(\lambda_m\).
That is to say, when we move from \(\lambda_{m-1}\) to \(\lambda_m\), we
first check this condition using the previously computed solution for
\(\hat{\vbeta}(\lambda_{m-1})\), and then perform blockwise coordinate
descent, using only those groups that failed this inequality.

This discarding rule is fast, because it uses the previously computed
solution combined with a simple inequality, and, in practice, it tends
to accurately discard large numbers of groups. However, we should
reiterate that it is possible for the strong rule to fail. The Lipschitz
assumption is not a guarantee. Because of this, it is critical that,
after discarding some of the groups and running the algorithm on the
others, the KKT condition is checked on all discarded groups. If there
are no violations, then we have the solution.

To minimize gradient computations for groups that will eventually be
determined to be inactive, we actually keep track of two sets: the
strong set \(\mathcal{S}\) and the active set \(\mathcal{A}\). The
active set collects all groups that have ever had non-zero coefficients
at previous values of \(\lambda\). We first iterate over previously
active groups, then check the strong set to see if we missed any, and
finally check all the remaining groups. When the number of groups is
very large, this avoids onerous computations for as many groups as
possible. The complete algorithm including this active set iteration is
shown in \autoref{alg:djm}.

\begin{algorithm}[tb!]
  \begin{algorithmic}[1]
  \STATE {\bfseries Input:} $\mX$, $\vY$, $\mathcal{G}$, and $\{\lambda_1,\ldots,\lambda_M\}$.\;\; \textbf{Output:} $\hat\vbeta$.
  \STATE {\bfseries Initialize:} $\mathcal{A} = \mathcal{S} = \varnothing$, $\hat\vbeta = 0$.
  \FOR{$m=1$ \TO  $M$} 
  \STATE \textbf{Update} $\mathcal{S} \leftarrow \mathcal{S} \bigcup \left\{
  g\in\mathcal{S}^c : \snorm{S(\nabla \ell(\hat\vbeta_g),\ t\alpha\lambda_m)}_2 > t(1-\alpha)\lambda_m \right\}$.
  \STATE \textbf{Apply} \autoref{alg:sparsegl} with $\mathcal{G} = \mathcal{A}$ (MM gradient update).
  \STATE \textbf{Update} $\mathcal{A} \leftarrow \mathcal{A} \bigcup \left\{g \in\mathcal{S}\bigcap\mathcal{A}^c : \snorm{S(\nabla \ell(\hat\vbeta_g),\ t\alpha\lambda_m)}_2 > t(1-\alpha)\lambda_m\right\}.$
  \begin{ALC@g}
  \STATE If there were any violations, \textbf{go to} to Line 5.
  \end{ALC@g}
  \STATE \textbf{Update} $\mathcal{A} \leftarrow \mathcal{A}\bigcup \left\{g\in\mathcal{S}^c\bigcap\mathcal{A}^c : \snorm{S(\nabla \ell(\hat\vbeta_g),\ t\alpha\lambda_m)}_2 > t(1-\alpha)\lambda_m\right\}.$
  \begin{ALC@g}
  \STATE If there were any violations, \textbf{go to} to Line 5.
  \end{ALC@g}
  \STATE \textbf{Set} $\mathcal{S} = \mathcal{S} \bigcup \mathcal{A}$.
  \ENDFOR
  \end{algorithmic}
  \caption{Sequential strong rule and Majorization Minimization in \pkg{sparsegl}}
  \label{alg:djm}
\end{algorithm}

\hypertarget{risk-estimation}{%
\subsection{Risk estimation}\label{risk-estimation}}

For many regularized prediction methods, tuning parameter selection is
largely performed with cross validation. However, cross validation can
be computationally expensive when the data set is large enough that the
initial fit is slow. For this reason, \pkg{sparsegl} provides
information criteria as well as cross validation routines.

In the Gaussian linear regression model given by
Equation~\ref{eq:linmod}, if \(\sigma\) is unknown then a general form
for a family of information criteria is given by \begin{equation}
  \label{eq:GICsigUnknown}
  \textrm{info}(C_n,g) 
  =\log\left(\frac{1}{n}\snorm{\vY-\mX\hat\vbeta}_2^2\right) +
  C_n \; g(\df),
\end{equation} where \(C_n\) depends only on \(n\),
\(g: [0,\infty) \rightarrow \mathbb{R}\) is a fixed function, and the
degrees of freedom (\(\df\)) measures the complexity of the estimation
procedure. The choices \(C_n = 2/n\) or \(C_n = \log(n)/n\) with
\(g(x) = x\) are commonly referred to as AIC \citep{Akaike1973} and BIC
\citep{Schwarz1978}, respectively. Additionally, generalized cross
validation \citep[GCV]{golub1979generalized} is defined as
\begin{equation*}
\textrm{GCV} = \frac{\frac{1}{n}\snorm{\vY-\mX\hat\vbeta}_2^2}{(1-\df/n)^2}.
\label{eq:gcv}
\end{equation*} Written on the log scale, GCV takes the form of
Equation~\ref{eq:GICsigUnknown} with \(g(x) = \log(1-x/n)\) and
\(C_n = -2\).

The key components for all three information criteria are the negative
log likelihood and the degrees of freedom. The first is simply a
function of the in-sample (training) error. On the other hand, the
degrees of freedom, while simple in the unregularized linear model (it
is the number of parameters), is less obvious for the sparse group
lasso. In general, the degress of freedom for any predictor
\(\hat{\vy}\) of \(\vY\) is defined as \citep{Efron1986}
\[\mathrm{df}(\hat\vy) = \frac{1}{\sigma^2}\sum_{i=1}^n\mathrm{Cov}(\vy_i,\ \hat\vy_i).\]
In the case of any linear predictor, \(\hat\vy = \mathbf{A}\vy\) for
some matrix \(\mathbf{A}\), it is easy to see that
\(\mathrm{df} = \mathrm{tr}(\mathbf{A})\). \citet{vaiter2012degrees}
gives an explicit formula for the group lasso (without intragroup
sparsity), but only minor modifications are required for the sparse
group lasso. We give a simplified version of their result here.

\begin{proposition}[\citealt{vaiter2012degrees}]
Suppose that for a fixed $\lambda >0$, the active set of $\hat{\beta}$ is $\mathcal{A}$ and that $\mX_{\mathcal{A}}$ is the set of columns associated to $\mathcal{A}$. Assume that $\mX_{\mathcal{A}}$ has full column rank. Then, 
$$\df = \mathrm{tr}\left(\mX_\mathcal{A}\left(\mX_\mathcal{A}^\T \mX_\mathcal{A} + \lambda \mathbf{K}\right)^{-1}\mX_{\mathcal{A}}^\T\right).$$
Here, $\mathbf{K}\in \R^{\mathcal{A}\times\mathcal{A}}$ is a block diagonal matrix with each block corresponding to a group $g$ having at least 1 nonzero $\hat\beta$. For such a group $g$, denote $\hat\beta_{g | \mathcal{A}}$ the subvector of nonzero coefficient estimates. Then $$\mathbf{K}_g = \frac{1}{\snorm{\hat\vbeta_{g | \mathcal{A}}}_2}\left(\mathbf{I} - \frac{\hat\vbeta_{g | \mathcal{A}} \hat\vbeta_{g | \mathcal{A}}^\T}{\snorm{\hat\vbeta_{g | \mathcal{A}}}^2_2} \right).$$
\end{proposition}

As long as the number of nonzero coefficients \(|\mathcal{A}|\) is
reasonably small, the degrees of freedom can be efficiently calculated
for each value of \(\lambda\). However, this calculation is generally
cubic in \(|\mathcal{A}|\). In these cases, an approximation may be
desired. We have found, in practice, that
\(\lambda \mathbf{K}\approx \mathbf{0}\) is reasonably accurate,
suggesting that \(\mathrm{df} \approx |\mathcal{A}|\) is also
reasonable. This approximation is exact for the lasso with \(\alpha=1\)
\citep{ZouHastie2007}.

\hypertarget{example-usage}{%
\section{Example usage}\label{example-usage}}

This section provides a simple illustration of using the \pkg{sparsegl}
package \citep{R-sparsegl} to fit the regularization path for sparse
group-lasso penalized learning problems. We first examine the linear
regression model when the response variable is continuous and then
briefly go over the logistic regression case. The package is available
from the Comprehensive R Archive Network (CRAN) at
\url{https://CRAN.R-project.org/package=sparsegl} and can be installed
and loaded in the usual manner:\footnote{The development version of the
  package is hosted at \url{https://github.com/dajmcdon/sparsegl} with
  accompanying
  \href{https://dajmcdon.github.io/sparsegl}{documentation}.}

\begin{Shaded}
\begin{Highlighting}[]
\FunctionTok{install.packages}\NormalTok{(}\StringTok{"sparsegl"}\NormalTok{)}
\FunctionTok{library}\NormalTok{(}\StringTok{"sparsegl"}\NormalTok{)}
\end{Highlighting}
\end{Shaded}

We first create a small simulated dataset along with a vector indicating
the grouping structure.

\begin{Shaded}
\begin{Highlighting}[]
\FunctionTok{set.seed}\NormalTok{(}\DecValTok{1010}\NormalTok{)}
\NormalTok{n }\OtherTok{\textless{}{-}} \DecValTok{100}
\NormalTok{p }\OtherTok{\textless{}{-}} \DecValTok{200}
\NormalTok{X }\OtherTok{\textless{}{-}} \FunctionTok{matrix}\NormalTok{(}\FunctionTok{rnorm}\NormalTok{(n}\SpecialCharTok{*}\NormalTok{p), }\AttributeTok{nrow =}\NormalTok{ n, }\AttributeTok{ncol =}\NormalTok{ p)}
\NormalTok{beta }\OtherTok{\textless{}{-}} \FunctionTok{c}\NormalTok{(}
  \FunctionTok{rep}\NormalTok{(}\DecValTok{5}\NormalTok{, }\DecValTok{5}\NormalTok{), }\FunctionTok{c}\NormalTok{(}\DecValTok{5}\NormalTok{, }\SpecialCharTok{{-}}\DecValTok{5}\NormalTok{, }\DecValTok{2}\NormalTok{, }\DecValTok{0}\NormalTok{, }\DecValTok{0}\NormalTok{), }\FunctionTok{rep}\NormalTok{(}\SpecialCharTok{{-}}\DecValTok{5}\NormalTok{, }\DecValTok{5}\NormalTok{), }\FunctionTok{c}\NormalTok{(}\DecValTok{2}\NormalTok{, }\SpecialCharTok{{-}}\DecValTok{3}\NormalTok{, }\DecValTok{8}\NormalTok{, }\DecValTok{0}\NormalTok{, }\DecValTok{0}\NormalTok{),}
  \FunctionTok{rep}\NormalTok{(}\DecValTok{0}\NormalTok{, (p }\SpecialCharTok{{-}} \DecValTok{20}\NormalTok{))}
\NormalTok{)}
\NormalTok{groups }\OtherTok{\textless{}{-}} \FunctionTok{rep}\NormalTok{(}\DecValTok{1}\SpecialCharTok{:}\NormalTok{(p }\SpecialCharTok{/} \DecValTok{5}\NormalTok{), }\AttributeTok{each =} \DecValTok{5}\NormalTok{)}
\NormalTok{eps }\OtherTok{\textless{}{-}} \FunctionTok{rnorm}\NormalTok{(n, }\AttributeTok{mean =} \DecValTok{0}\NormalTok{, }\AttributeTok{sd =} \DecValTok{1}\NormalTok{)}
\NormalTok{y }\OtherTok{\textless{}{-}}\NormalTok{ X }\SpecialCharTok{\%*\%}\NormalTok{ beta }\SpecialCharTok{+}\NormalTok{ eps}
\NormalTok{pr }\OtherTok{\textless{}{-}} \DecValTok{1} \SpecialCharTok{/}\NormalTok{ (}\DecValTok{1} \SpecialCharTok{+} \FunctionTok{exp}\NormalTok{(}\SpecialCharTok{{-}}\NormalTok{X }\SpecialCharTok{\%*\%}\NormalTok{ beta))}
\NormalTok{y0 }\OtherTok{\textless{}{-}} \FunctionTok{rbinom}\NormalTok{(n, }\DecValTok{1}\NormalTok{, pr)}
\end{Highlighting}
\end{Shaded}

The \pkg{sparsegl} package is mainly used with calls to 2 functions:

\begin{itemize}
\tightlist
\item
  \texttt{sparsegl()}: fits sparse group regularized regression and
  classification models;
\item
  \texttt{cv.sparsegl()}: repeatedly calls \texttt{sparsegl()} for the
  purposes of tuning parameter selection via cross validation.
\end{itemize}

The interface is intended to closely mimic that available in other
\proglang{R} packages for regularized linear models, most notably
\pkg{glmnet} \citep{R-glmnet}. To perform the regularization path
fitting at a sequence of regularization parameters, \texttt{sparsegl()}
takes as required inputs, only \texttt{x}, the design matrix, and
\texttt{y}, the response vector. Other optional arguments are the
grouping vector \texttt{group}, the \texttt{family} (either
\texttt{"gaussian"} or \texttt{"binomial"}), a penalty vector for group
weights other than the size, the relative weight of lasso penalty,
desired lower or upper bounds for coefficient estimates, and other
optional configurations.

\begin{Shaded}
\begin{Highlighting}[]
\NormalTok{fit }\OtherTok{\textless{}{-}} \FunctionTok{sparsegl}\NormalTok{(X, y, }\AttributeTok{group =}\NormalTok{ groups)}
\end{Highlighting}
\end{Shaded}

\begin{figure}[t!]

{\centering \includegraphics[width=0.45\linewidth]{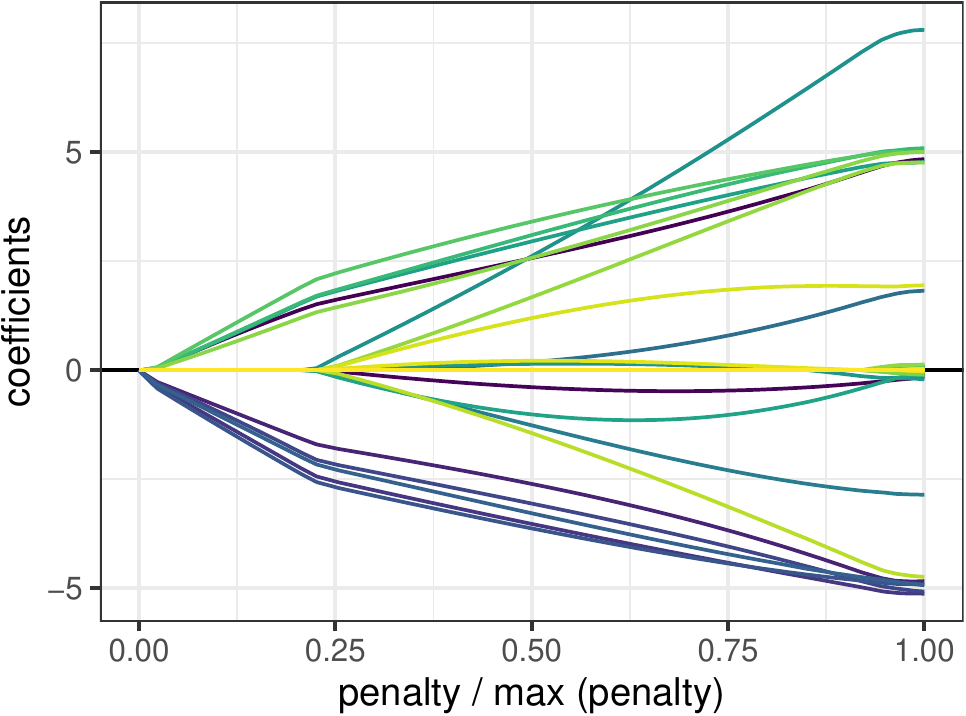} \includegraphics[width=0.45\linewidth]{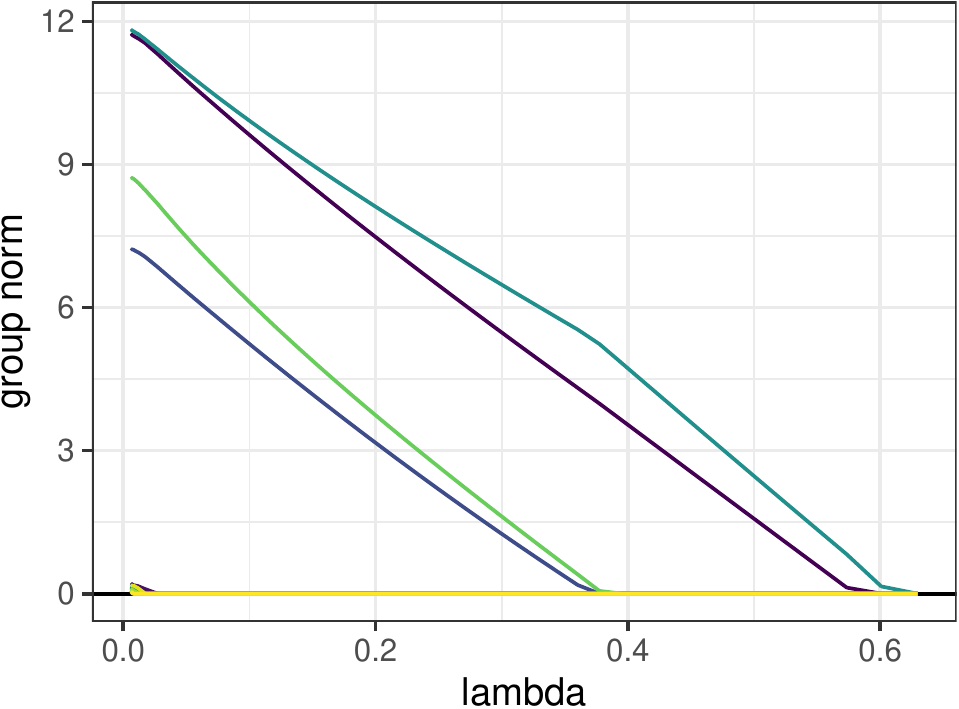} 

}

\caption{The left panel plots the estimated coefficients aganist the sparse group penalty; while the right plots the $\ell_2$-norm of each group against $\lambda$.}\label{fig:coef-trace}
\end{figure}

We include a number of \pkg{S3} methods for \texttt{sparsegl} typical
for linear models: \texttt{plot()}, \texttt{coef()}, \texttt{predict()}
and \texttt{print()}. The \texttt{plot()} function displays either the
coefficients or the group norms on the \(y\)-axis against either
\(\{\lambda_m\}_{m=1}^M\) or the scaled penalty on the
\(x\)-axis.\footnote{We have chosen to implement plotting throughout the
  package using \pkg{ggplot2} \citep{R-ggplot2}.} The resulting figures
are shown in \autoref{fig:coef-trace}.

\begin{Shaded}
\begin{Highlighting}[]
\FunctionTok{plot}\NormalTok{(fit, }\AttributeTok{y\_axis =} \StringTok{"coef"}\NormalTok{, }\AttributeTok{x\_axis =} \StringTok{"penalty"}\NormalTok{, }\AttributeTok{add\_legend =} \ConstantTok{FALSE}\NormalTok{)}
\FunctionTok{plot}\NormalTok{(fit, }\AttributeTok{y\_axis =} \StringTok{"group"}\NormalTok{, }\AttributeTok{x\_axis =} \StringTok{"lambda"}\NormalTok{, }\AttributeTok{add\_legend =} \ConstantTok{FALSE}\NormalTok{)}
\end{Highlighting}
\end{Shaded}

The \texttt{coef()} and \texttt{predict()} methods give the coefficients
or predicted values for a new design matrix \(\widetilde{\mX}\) at the
requested \(\lambda\)'s, potentially allowing for \(\lambda\) values
different from those used at the fitting stage.

\begin{Shaded}
\begin{Highlighting}[]
\FunctionTok{coef}\NormalTok{(fit, }\AttributeTok{s =} \FunctionTok{c}\NormalTok{(}\FloatTok{0.02}\NormalTok{, }\FloatTok{0.03}\NormalTok{))[}\FunctionTok{c}\NormalTok{(}\DecValTok{1}\NormalTok{, }\DecValTok{3}\NormalTok{, }\DecValTok{25}\NormalTok{, }\DecValTok{29}\NormalTok{), ]}
\end{Highlighting}
\end{Shaded}

\begin{verbatim}
## 4 x 2 sparse Matrix of class "dgCMatrix"
##                      s1         s2
## (Intercept) -0.05189536 -0.1718156
## V2           4.71082485  4.6339817
## V24          .           .        
## V28          .           .
\end{verbatim}

\begin{Shaded}
\begin{Highlighting}[]
\FunctionTok{predict}\NormalTok{(fit, }\AttributeTok{newx =} \FunctionTok{tail}\NormalTok{(X), }\AttributeTok{s =}\NormalTok{ fit}\SpecialCharTok{$}\NormalTok{lambda[}\DecValTok{2}\SpecialCharTok{:}\DecValTok{3}\NormalTok{])}
\end{Highlighting}
\end{Shaded}

\begin{verbatim}
##               s1        s2
##  [95,] -3.894658 -2.966973
##  [96,] -3.906349 -2.945468
##  [97,] -4.119689 -4.241786
##  [98,] -4.184564 -4.555082
##  [99,] -4.175593 -4.382721
## [100,] -4.071804 -4.091689
\end{verbatim}

\begin{Shaded}
\begin{Highlighting}[]
\FunctionTok{print}\NormalTok{(fit)}
\end{Highlighting}
\end{Shaded}

\begin{verbatim}
## 
## Call:  sparsegl(x = X, y = y, group = groups) 
## 
## Summary of Lambda sequence:
##          lambda index nnzero active_grps
## Max.    0.62948     1      0           0
## 3rd Qu. 0.19676    26     20           4
## Median  0.06443    50     19           4
## 1st Qu. 0.02014    75     25           5
## Min.    0.00629   100    111          23
\end{verbatim}

\begin{figure}[t!]

{\centering \includegraphics{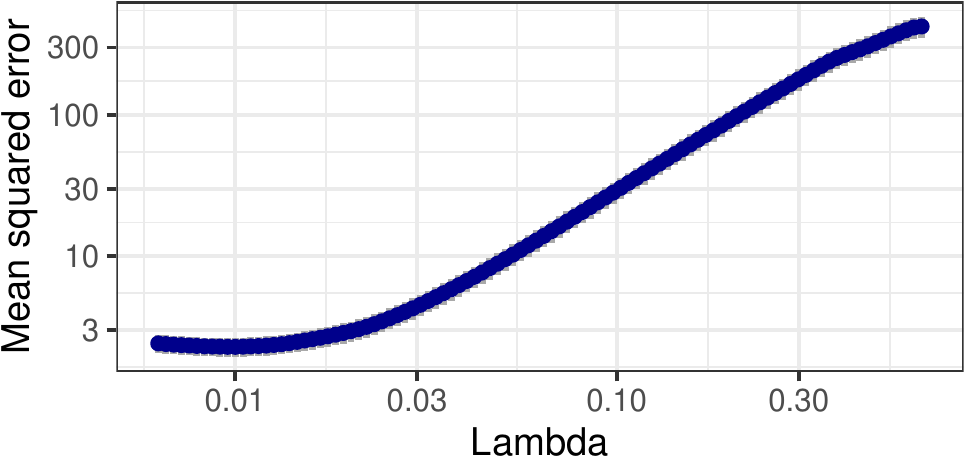} 

}

\caption{The cross validation estimate of out-of-sample prediction mean-squared error is displayed against the sequence of $\lambda$ values.}\label{fig:cv-plot}
\end{figure}

The \texttt{cv.sparsegl()} function implements \(K\)-fold cross
validation and has a similar signature to \texttt{sparsegl()}. It allows
the user to choose the number of splits and the loss function for
measuring prediction or classification accuracy on the held-out sets.
Here, the \pkg{S3} \texttt{plot()} method displays the cross-validation
curve with upper and lower confidence bounds calculated as \(\pm 1\)
standard error across the folds for each \(\lambda\) in the
regularization path (\autoref{fig:cv-plot}).

\begin{Shaded}
\begin{Highlighting}[]
\NormalTok{cv\_fit }\OtherTok{\textless{}{-}} \FunctionTok{cv.sparsegl}\NormalTok{(X, y, groups, }\AttributeTok{nfolds =} \DecValTok{15}\NormalTok{)}
\FunctionTok{plot}\NormalTok{(cv\_fit)}
\end{Highlighting}
\end{Shaded}

The \texttt{coef()} and \texttt{predict()} methods work similarly to
those above. The only difference being that they can additionally accept
the strings \texttt{lambda.min} or \texttt{lambda.1se}, respectively the
\(\lambda\) that minimizes the average cross validation error and the
largest \(\lambda\) such that the cross-validated prediction error is
within 1 standard error of the minimum.

\begin{Shaded}
\begin{Highlighting}[]
\FunctionTok{coef}\NormalTok{(cv\_fit, }\AttributeTok{s =} \StringTok{"lambda.1se"}\NormalTok{)[}\FunctionTok{c}\NormalTok{(}\DecValTok{1}\NormalTok{, }\DecValTok{3}\NormalTok{, }\DecValTok{25}\NormalTok{, }\DecValTok{29}\NormalTok{), ]}
\end{Highlighting}
\end{Shaded}

\begin{verbatim}
## (Intercept)          V2         V24         V28 
## 0.004435981 4.740139458 0.000000000 0.000000000
\end{verbatim}

\begin{Shaded}
\begin{Highlighting}[]
\FunctionTok{predict}\NormalTok{(cv\_fit, }\AttributeTok{newx =} \FunctionTok{tail}\NormalTok{(X), }\AttributeTok{s =} \StringTok{"lambda.min"}\NormalTok{) }\SpecialCharTok{|\textgreater{}} \FunctionTok{c}\NormalTok{()}
\end{Highlighting}
\end{Shaded}

\begin{verbatim}
## [1]  11.364725  39.985246   4.635314 -34.832413  -6.602096 -16.138344
\end{verbatim}

\begin{figure}[t!]

{\centering \includegraphics{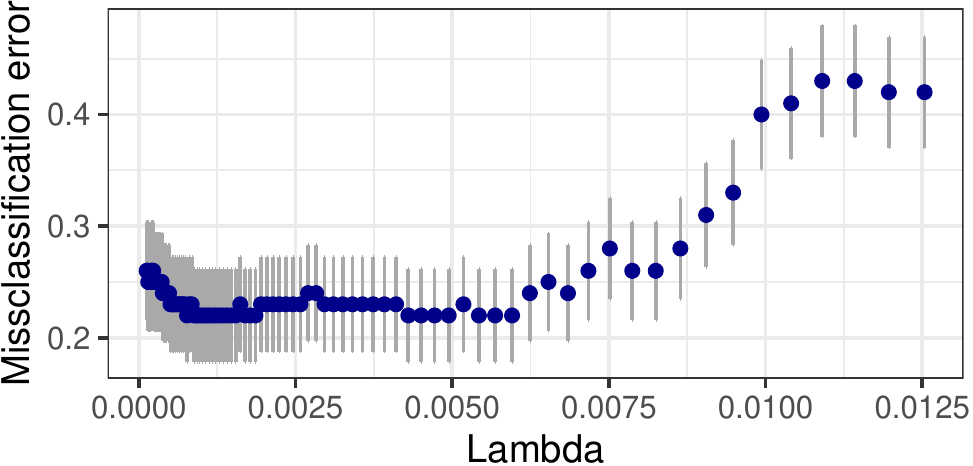} 

}

\caption{The cross validation estimate of out-of-sample error is displayed against the sequence of $\lambda$ values. For logistic regression, misclassification error may be used.}\label{fig:logitres}
\end{figure}

For logistic regression, only a different \texttt{family} is required.
Cross validation can be implemented with misclassification or deviance
loss (\autoref{fig:logitres}).

\begin{Shaded}
\begin{Highlighting}[]
\NormalTok{fit\_logit }\OtherTok{\textless{}{-}} \FunctionTok{sparsegl}\NormalTok{(X, y0, groups, }\AttributeTok{family =} \StringTok{"binomial"}\NormalTok{)}
\NormalTok{cv\_fit\_logit }\OtherTok{\textless{}{-}} \FunctionTok{cv.sparsegl}\NormalTok{(}
\NormalTok{  X, y0, groups, }\AttributeTok{family =} \StringTok{"binomial"}\NormalTok{, }\AttributeTok{pred.loss =} \StringTok{"misclass"}
\NormalTok{)}
\FunctionTok{plot}\NormalTok{(cv\_fit\_logit, }\AttributeTok{log\_axis =} \StringTok{"none"}\NormalTok{)}
\end{Highlighting}
\end{Shaded}

In some cases, when computations are at a premium, cross validation my
be too demanding for the purposes of risk estimation. For this reason,
\pkg{sparsegl} provides an \texttt{estimate\_risk()} function. It can be
used to compute any of AIC \citep{Akaike1973}, BIC \citep{Schwarz1978},
and GCV \citep{golub1979generalized}. All three are computed by
combining the log-likelihood with a penalty term for model flexibility.
In addition to a fitted \texttt{sparsegl} model,
\texttt{estimate\_risk()} also needs the original design matrix. Because
the exact degrees-of-freedom can be computationally expensive, setting
\texttt{approx\_df\ =\ TRUE} uses the number of non-zero coefficient
estimates, which can be reasonably accurate.

\begin{Shaded}
\begin{Highlighting}[]
\NormalTok{er }\OtherTok{\textless{}{-}} \FunctionTok{estimate\_risk}\NormalTok{(fit, X)}
\end{Highlighting}
\end{Shaded}

\begin{figure}[t!]

{\centering \includegraphics{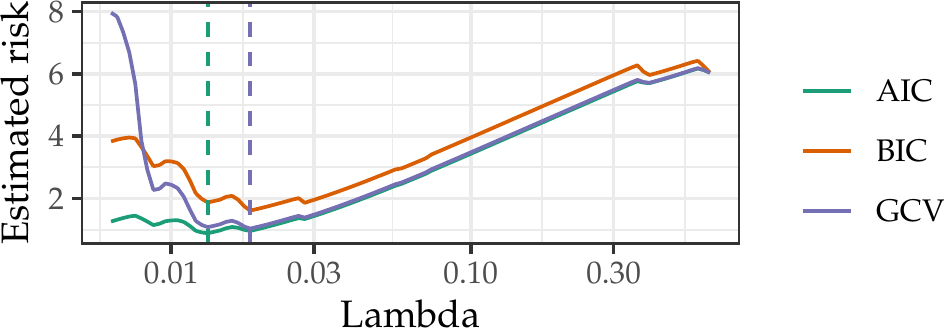} 

}

\caption{For Gaussian loss, AIC, BIC, and GCV (solid lines) along with their minima (vertical dashed lines) can be estimated.}\label{fig:re-plot}
\end{figure}

In this simulation, the \(\lambda\) that minimizes AIC is 0.013 while
the CV minimizer is 0.01. The estimated risk curves are plotted against
\(\lambda\) in \autoref{fig:re-plot}.

Additional documentation and examples are provided on the package
\href{https://dajmcdon.github.io/sparsegl}{website}.

\hypertarget{applications}{%
\section{Applications}\label{applications}}

We examine two applications for which sparse group lasso is a natural
estimator. The first uses data regarding COVID-19 and trustworthiness of
information sources, which is included in the package. The second uses a
very large though sparse data set from neuroimaging. Finally, we briefly
investigate the accuracy of \pkg{sparsegl} relative to \pkg{gglasso} and
\pkg{CVXR}.

\hypertarget{sec:trust}{%
\subsection{Geographic distribution of trust in
experts}\label{sec:trust}}

Two typical uses for sparse group lasso are (1) additive models where
continuous predictors are expanded in a basis and (2) discrete factors
as predictors. Here we demonstrate an example using both at the same
time. We examine data from The Delphi Group at Carnegie Mellon
University U.S.
\href{https://cmu-delphi.github.io/delphi-epidata/symptom-survey/contingency-tables.html}{COVID-19
Trends and Impact Survey (CTIS)}, in partnership with Facebook. In
particular, we examine the publicly available contingency table reports,
which break down survey responses by age, race/ethnicity, gender, and
other demographic variables of interest. The necessary data to reproduce
this analysis is included in \pkg{sparsegl} as \texttt{trust\_experts}.

In particular, we will focus on the ``estimated percentage of
respondents who trust \(\ldots\) to provide accurate news and
information about COVID-19.'' This survey item is reported for a variety
of different potential sources of information---personal doctors/nurses,
scientists, the Centers for Disease Control (CDC), government health
officials, politicians, journalists, friends, and religious leaders. In
this analysis, we average the first 4, characterize the combination as
``experts'', and use this as the response variable in a linear model.

We regress ``trust in experts'' on 5 factor predictors representing
month of report, state of residence, age group, race/ethnicity, and
self-reported gender identity. We also include two continuous
predictors: ``estimated percentage of people with Covid-like illness''
and ``estimated percentage of people reporting Covid-like illness in
their local community, including their household'' to control for the
amount of exposure that respondents may have been having to the
pandemic. Both continuous predictors are incorporated with a B-spline
basis expansion and 10 degrees-of-freedom. The result is largely similar
to a generalized additive model as implemented with \pkg{mgcv}
\citep{woodGams,R-mgcv}. The design matrix can be created using the
following code:

\begin{Shaded}
\begin{Highlighting}[]
\FunctionTok{library}\NormalTok{(}\StringTok{"dplyr"}\NormalTok{)}
\FunctionTok{library}\NormalTok{(}\StringTok{"splines"}\NormalTok{)}
\FunctionTok{data}\NormalTok{(}\StringTok{"trust\_experts"}\NormalTok{, }\AttributeTok{package =} \StringTok{"sparsegl"}\NormalTok{)}
\NormalTok{trust\_experts }\OtherTok{\textless{}{-}} \FunctionTok{mutate}\NormalTok{(trust\_experts, }\FunctionTok{across}\NormalTok{(}
  \FunctionTok{where}\NormalTok{(is.factor), }
  \SpecialCharTok{\textasciitilde{}} \StringTok{\textasciigrave{}}\AttributeTok{attr\textless{}{-}}\StringTok{\textasciigrave{}}\NormalTok{(.x, }\StringTok{"contrasts"}\NormalTok{, }\FunctionTok{contr.sum}\NormalTok{(}\FunctionTok{nlevels}\NormalTok{(.x), }\AttributeTok{contrasts =} \ConstantTok{FALSE}\NormalTok{))}
\NormalTok{))}
\NormalTok{x }\OtherTok{\textless{}{-}}\NormalTok{ Matrix}\SpecialCharTok{::}\FunctionTok{sparse.model.matrix}\NormalTok{(}
\NormalTok{  trust\_experts }\SpecialCharTok{\textasciitilde{}} \DecValTok{0} \SpecialCharTok{+}\NormalTok{ region }\SpecialCharTok{+}\NormalTok{ age }\SpecialCharTok{+}\NormalTok{ gender }\SpecialCharTok{+}\NormalTok{ raceethnicity }\SpecialCharTok{+}\NormalTok{ period }\SpecialCharTok{+}
    \FunctionTok{bs}\NormalTok{(cli, }\AttributeTok{df =} \DecValTok{10}\NormalTok{) }\SpecialCharTok{+} \FunctionTok{bs}\NormalTok{(hh\_cmnty\_cli, }\AttributeTok{df =} \DecValTok{10}\NormalTok{),}
  \AttributeTok{data =}\NormalTok{ trust\_experts, }\AttributeTok{drop.unused.levels =} \ConstantTok{TRUE}\NormalTok{)}
\end{Highlighting}
\end{Shaded}

\begin{figure}[t!]

{\centering \includegraphics[width=5in]{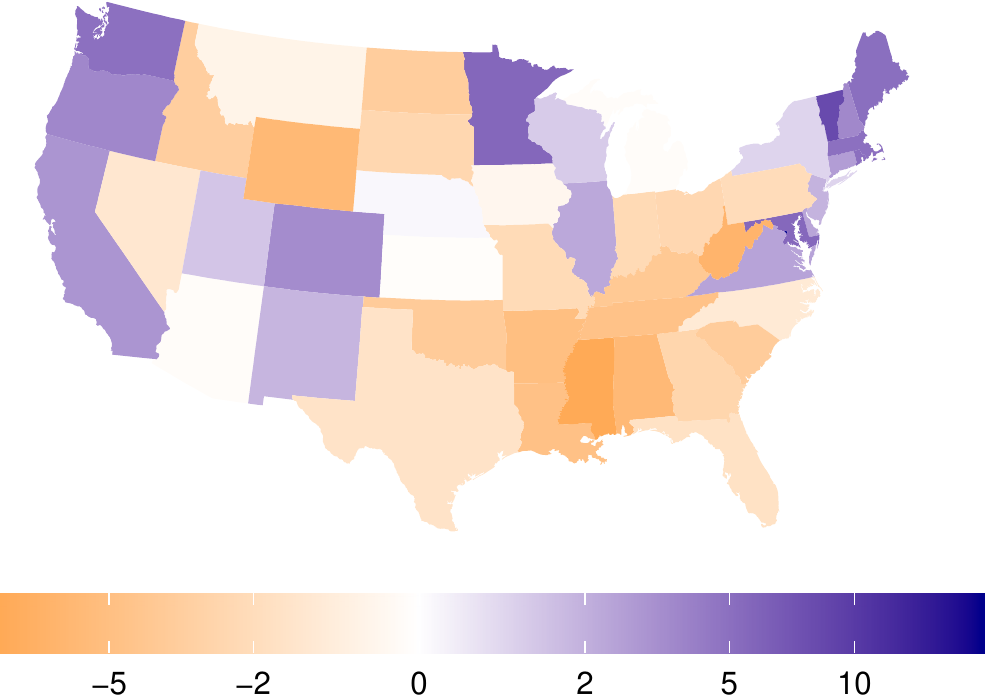} 

}

\caption{This figure displays estimates for each state's level of trust in experts' advice about Covid-19. The value displayed represents the change relative the U.S.-wide average.}\label{fig:trust}
\end{figure}

After omitting both structural and otherwise missing data, the final
model is estimated with 9759 observations on 101 predictors. As shown in
the code, we did not use contrasts, fully expanding each factor in a
one-hot encoding. This allows all estimated coefficients to be
interpreted as deviations from the grand mean conditional on continuous
predictors, which is natural. Such a formulation (along with the group
penalty) is closely related to Bayesian linear models with separate
Gaussian priors centered at 0 for each level of the factor. Other
contrasts could be used by modifying the above, but the interpretation
is more complicated. Encoded as a sparse matrix, this requires about 2.1
MB of RAM to store, as opposed to 8.5 MB if it were dense. We estimated
the model using \texttt{cv.sparsegl()} and default arguments. Finally,
we chose \(\lambda\) to be the largest lambda within 1 standard error of
the CV minimum (\texttt{lambda.1se}), resulting in a sparser model.
\autoref{fig:trust} displays the estimated coefficients for the
state-of-residence predictors. Even controlling for age, race, gender,
and the amount of circulating Covid-like illness, the United States
displays strong geographic disparities when it comes to citizens' trust
in scientists and other health authorities.

\hypertarget{estimating-white-matter-connectivity}{%
\subsection{Estimating white matter
connectivity}\label{estimating-white-matter-connectivity}}

\begin{figure}[t!]

{\centering \includegraphics[width=0.75\linewidth]{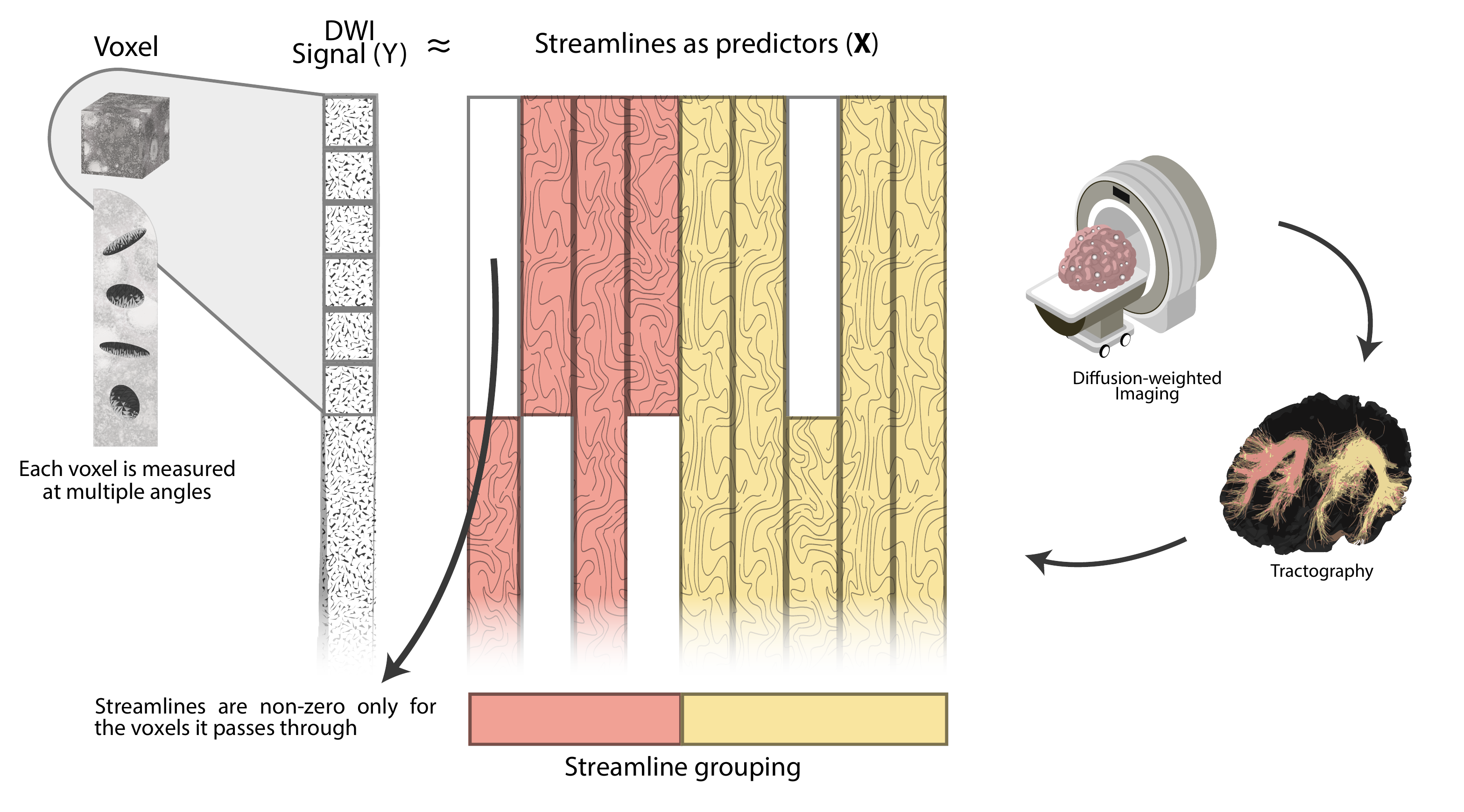} 

}

\caption{This graphic illustrates how the streamlines and voxels are converted from a diffusion-weighted image to a linear model. Each voxel is measured on 90 angles, so it occupies 90 rows in the data. When a streamline (column of $\mathbf{X}$) passes through a voxel, the values within that voxel are given by a physical model based on the direction of passage. Otherwise, if the streamline does not cross the voxel, the respective rows are zero.}\label{fig:dwi-model-description}
\end{figure}

\citet{pestilli2014} formulated an optimization model that takes as
input a set of brain connections generated using tractography algorithms
and predicts the MRI diffusion signal via a linear model
\citep{pestilli2014, daducci2015}. The \citet{pestilli2014} model had no
regularization, but \citet{aminmansour2019} extended the problem to
include group-regularization (this is an approach recently followed up
by \citealp{Schiavi2020}). In this study, we re-implemented the
\citet{aminmansour2019} formulation using \pkg{sparsegl} to illustrate
the feasibility and efficiency of the DWI modeling.

The neurological model predicts the DWI signal using the tractogram,
apportioning the image signal at each voxel to the streamlines according
to the measured gradient field. This preprocessing is shown pictorially
\autoref{fig:dwi-model-description}. We estimate streamline weights
using sparse group lasso, allowing the amount of regularization applied
to each group to be proportional to their cardinality. For our study, we
used one subject from the Human Connectome Project \citep{vanessen2012}.
The full brain tractogram has 3M streamlines, though we used only the
streamlines identified as being part of the Arcuate Fasciculus for
illustration.\footnote{The processed data used to estimate the sparse
  group lasso is available at
  \url{https://doi.org/10.6084/m9.figshare.20314917}.}

\citet{aminmansour2019} used an algorithm based on Orthogonal Matching
Pursuit to estimate a related model. The data used in that study
measures diffusion in 11,823 voxels using 96 magnetic field angles and
attempts to reconstruct the image using the ENCODE method
\citep{encode2017}, resulting in 1057 orientations and 868 fascicles.
This results in a linear regression problem with \(n\approx1M\) and
\(p = 868\). In comparison, our data contains 77,630 voxels measured on
90 angles, with a tractography of 10,244 streamlines. The resulting
linear model has \(n\approx6.9M\) and \(p\approx88K\), around
\(700\times\) the size of the previous analysis. The design matrix would
occupy over 500 GB if it were dense, but since it is only about 0.02\%
non-sparse, it requires 1.8 GB of memory when stored in CSC format.

\begin{figure}[t!]

{\centering \includegraphics{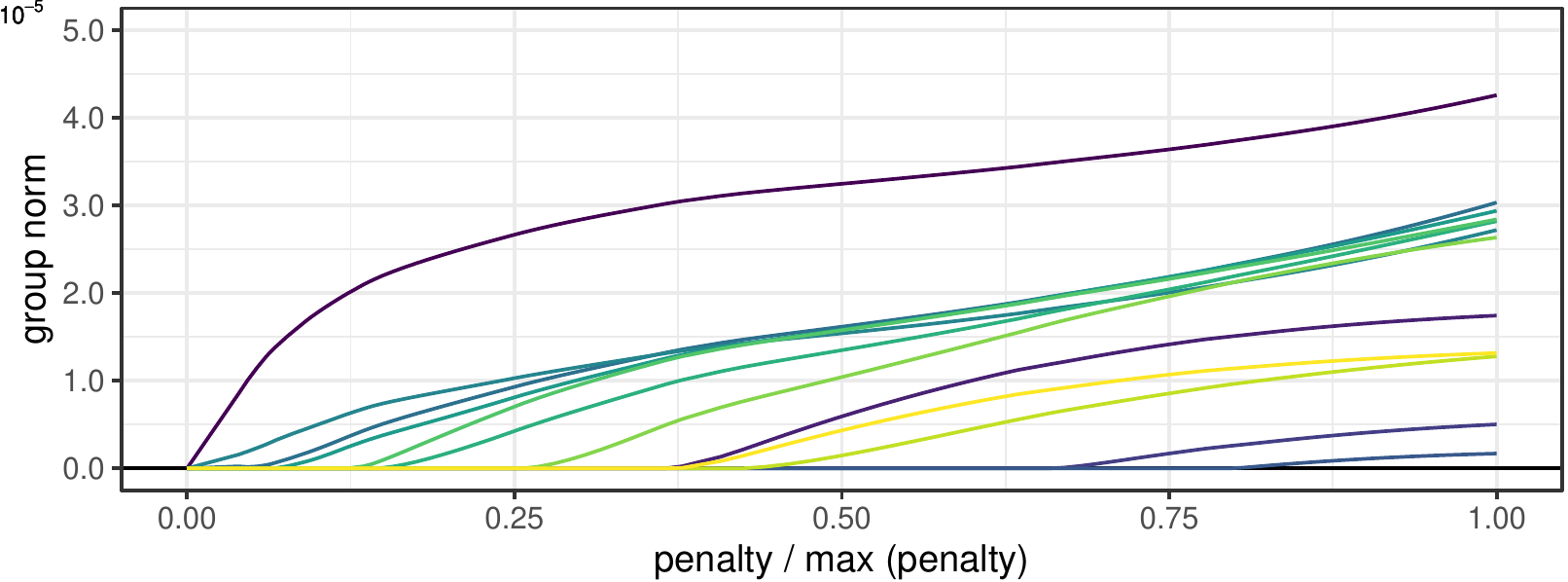} 

}

\caption{The group norm of the 12 groups based on neuroanatomical structure is plotted against the magnitude of the penalty.}\label{fig:dwi-fit}
\end{figure}

Estimating the group lasso using \pkg{sparsegl} with 12 groups and 100
values of \(\lambda\) required a little over 1 minute and about 6 GB of
peak memory usage on an Intel i7-11700K PC with 64 GB of RAM. The
previous method required nearly a day for a single value of \(\lambda\).
\autoref{fig:dwi-fit} displays the group norms of the 12 groups against
the magnitude of the penalty.

\hypertarget{accuracy-on-synthetic-problems}{%
\subsection{Accuracy on synthetic
problems}\label{accuracy-on-synthetic-problems}}

\begin{figure}[t!]

{\centering \includegraphics{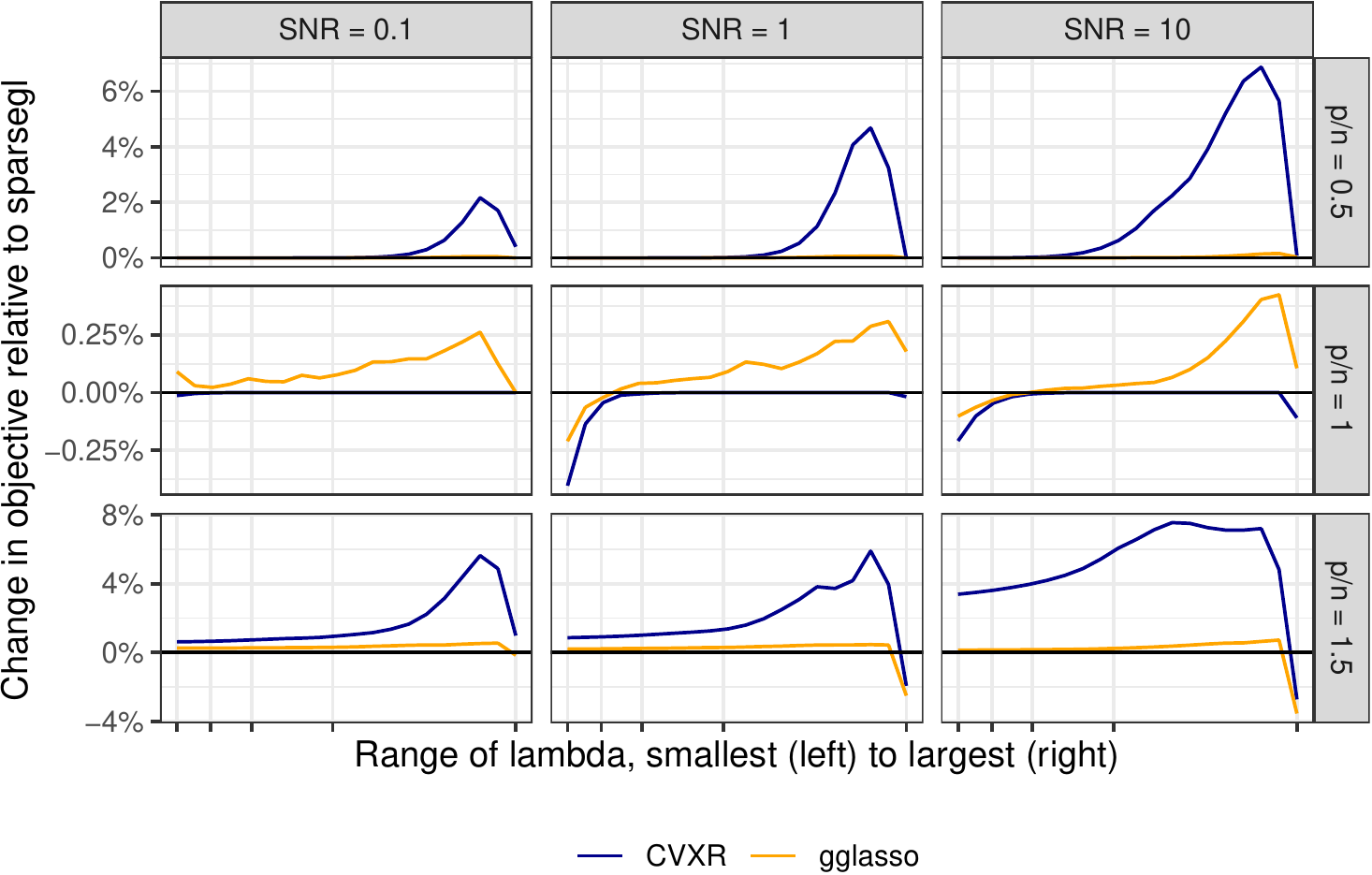} 

}

\caption{This figure shows the \% change in the objective function for \pkg{gglasso} and \pkg{CVXR} relative to \pkg{sparsegl}. The x-axis is scaled to make the $\lambda$ range comparable across conditions.}\label{fig:accuracy}
\end{figure}

A small-scale simulated comparison illustrates that \pkg{sparsegl} is
highly accurate in many regimes of interest. We generate synthetic data
from a linear model and examine the objective function for the estimated
model across a range of \(\lambda\). Specifically, we first generate the
predictors \(\mathbf{X}\) by simulating each element \(x_{ij}\) i.i.d.
standard Gaussian. We generate \(n=100\) observations and
\(p = \{50,\ 100,\ 150\}\) predictors. We use 10 groups in all cases,
with \[\beta = (\underbrace{1, \ldots, 1,}_{\textrm{group } 1}\ 
\underbrace{0, \dots, 0}_{\textrm{group } 2},\  
\underbrace{1, \ldots, 1}_{\textrm{group } 3},\ \ldots,\ 
\underbrace{0, \ldots, 0}_{\textrm{group } 10}).\] The groups are all of
size \(p/10\). The expected signal is then equal to \(p / 2\). We
simulate the response from Equation \ref{eq:linmod} with \(\sigma\)
chosen to produce expected signal-to-noise ratios (SNR) of
\(\{0.1,\ 1,\ 10\}\). We note that this design produces group sparsity,
even-numbered groups have no effect on the response, but it does not
produce within-group sparsity. \autoref{fig:accuracy} shows the results
for the 9 combinations of these conditions across 20 values of
\(\lambda\) determined automatically by \pkg{sparsegl} and reused for
the other packages. Note that \pkg{gglasso} is optimizing a different
objective function than \pkg{CVXR} or \pkg{sparsegl} which both use
\(\alpha = 0.2\). Despite this discrepancy, the objective values for
\pkg{gglasso} and \pkg{sparsegl} are generally quite close, with
\pkg{gglasso} tending to be slightly higher, as expected, but by no more
than 0.75\%. On the other hand \pkg{CVXR} is often much less accurate,
especially for large values of \(\lambda\). This divergence is due to
its inability to recover exact 0 solutions, tending to instead produce
estimates nearly, but not exactly, equal to 0. When \(p=n=100\),
\pkg{CVXR} is actually slightly more accurate, especially for large or
small \(\lambda\). Inspecting the estimated \(\widehat{\beta}\) in this
setting, it seems to have slightly less bias on the non-zero groups,
producing estimates with slightly larger magnitude.

\hypertarget{discussion}{%
\section{Discussion}\label{discussion}}

We developed this package for solving sparse group lasso optimization
problems using group \(\ell_2\) and \(\ell_1\) penalties with an eye
toward computational efficiency for very large, potentially sparse
design matrices. This efficiency is achieved through a customized
\proglang{Fortran} implementation, avoidance of deep copy behaviour, and
the use of sequential strong rules for the regularization parameter. We
also provide heuristics for tuning parameter selection without the need
for refitting inherent in cross-validation and enable some simple
extensions such as differential weights in the \(\ell_1\) penalty and
boundary constraints on the coefficients.

\hypertarget{acknowledgements}{%
\section*{Acknowledgements}\label{acknowledgements}}
\addcontentsline{toc}{section}{Acknowledgements}

In addition to the \proglang{R} packages described above, this
manuscript and \pkg{sparsegl} benefitted from \pkg{devtools}
\citep{R-devtools}, \pkg{usethis} \citep{R-usethis}, \pkg{testthat}
\citep{R-testthat}, \pkg{knitr} \citep{R-knitr}, and \pkg{rticles}
\citep{R-rticles}.

Franco Pestilli was supported with grants from the National Science
Foundation (OAC-1916518, IIS-1912270, IIS-1636893, BCS-1734853),
National Institutes of Health (NIMH R01MH126699, NIBIB R01EB030896,
NIBIB R01EB029272), a Microsoft Investigator Fellowship, and a gift from
the Kavli Foundation. Daniel J. McDonald was partially supported by the
National Science Foundation (DMS--1753171) and the National Sciences and
Engineering Research Council of Canada (RGPIN-2021-02618). The analysis
in \autoref{sec:trust} is based on survey results from Carnegie Mellon
University's Delphi Group.

\bibliographystyle{rss}
\bibliography{sparsegl.bib}

\begin{thebibliography}{45}
\expandafter\ifx\csname natexlab\endcsname\relax\def\natexlab#1{#1}\fi
\expandafter\ifx\csname url\endcsname\relax
  \def\url#1{\texttt{#1}}\fi
\expandafter\ifx\csname urlprefix\endcsname\relax\def\urlprefix{URL: }\fi

\bibitem[{Agrawal et~al.(2018)Agrawal, Verschueren, Diamond and
  Boyd}]{cvxpy-jcd}
Agrawal, A., Verschueren, R., Diamond, S. and Boyd, S. (2018) A rewriting
  system for convex optimization problems.
\newblock \textit{Journal of Control and Decision}, \textbf{5}, 42--60.

\bibitem[{Akaike(1973)}]{Akaike1973}
Akaike, H. (1973) Information theory and an extension of the maximum likelihood
  principle.
\newblock In \textit{Proceedings of the 2nd International Symposium of
  Information Theory} (eds. B.~N. Petrov and F.~Csaki), 267--281.

\bibitem[{Allaire et~al.(2022)Allaire, Xie, Dervieux, {R Foundation}, Wickham,
  {Journal of Statistical Software}, Vaidyanathan, {Association for Computing
  Machinery}, Boettiger, {Elsevier}, Broman, Mueller, Quast, Pruim, Marwick,
  Wickham, Keyes, Yu, Emaasit, Onkelinx, Gasparini, Desautels, Leutnant,
  {MDPI}, {Taylor and Francis}, Öğreden, Hance, Nüst, Uvesten, Campitelli,
  Muschelli, Hayes, Kamvar, Ross, Cannoodt, Luguern, Kaplan, Kreutzer, Wang,
  Hesselberth and Hyndman}]{R-rticles}
Allaire, J., Xie, Y., Dervieux, C., {R Foundation}, Wickham, H., {Journal of
  Statistical Software}, Vaidyanathan, R., {Association for Computing
  Machinery}, Boettiger, C., {Elsevier}, Broman, K., Mueller, K., Quast, B.,
  Pruim, R., Marwick, B., Wickham, C., Keyes, O., Yu, M., Emaasit, D.,
  Onkelinx, T., Gasparini, A., Desautels, M.-A., Leutnant, D., {MDPI}, {Taylor
  and Francis}, Öğreden, O., Hance, D., Nüst, D., Uvesten, P., Campitelli,
  E., Muschelli, J., Hayes, A., Kamvar, Z.~N., Ross, N., Cannoodt, R., Luguern,
  D., Kaplan, D.~M., Kreutzer, S., Wang, S., Hesselberth, J. and Hyndman, R.
  (2022) \textit{\pkg{rticles}: Article Formats for \proglang{R}
  \proglang{Markdown}}.
\newblock \urlprefix\url{https://CRAN.R-project.org/package=rticles}.
\newblock \proglang{R}~package version~0.24.

\bibitem[{Aminmansour et~al.(2019)Aminmansour, Patterson, Le, Peng, Mitchell,
  Pestilli, Caiafa, Greiner and White}]{aminmansour2019}
Aminmansour, F., Patterson, A., Le, L., Peng, Y., Mitchell, D., Pestilli, F.,
  Caiafa, C.~F., Greiner, R. and White, M. (2019) Learning macroscopic brain
  connectomes via group-sparse factorization.
\newblock In \textit{Advances in Neural Information Processing Systems} (eds.
  H.~Wallach, H.~Larochelle, A.~Beygelzimer, F.~d\textquotesingle
  Alch\'{e}-Buc, E.~Fox and R.~Garnett), vol.~32.
\newblock
  \urlprefix\url{https://proceedings.neurips.cc/paper/2019/hash/0bfce127947574733b19da0f30739fcd-Abstract.html}.

\bibitem[{Bache and Wickham(2022)}]{R-magrittr}
Bache, S.~M. and Wickham, H. (2022) \textit{\pkg{magrittr}: A Forward-Pipe
  Operator for \proglang{R}}.
\newblock \urlprefix\url{https://CRAN.R-project.org/package=magrittr}.
\newblock \proglang{R}~package version~2.0.3.

\bibitem[{Caiafa et~al.(2017)Caiafa, Sporns, Saykin and Pestilli}]{encode2017}
Caiafa, C.~F., Sporns, O., Saykin, A. and Pestilli, F. (2017) Unified
  representation of tractography and diffusion-weighted mri data using sparse
  multidimensional arrays.
\newblock In \textit{Advances in Neural Information Processing Systems} (eds.
  I.~Guyon, U.~V. Luxburg, S.~Bengio, H.~Wallach, R.~Fergus, S.~Vishwanathan
  and R.~Garnett), vol.~30.
\newblock
  \urlprefix\url{https://proceedings.neurips.cc/paper/2017/hash/ccbd8ca962b80445df1f7f38c57759f0-Abstract.html}.

\bibitem[{Candes and Tao(2007)}]{CandesTao2007}
Candes, E.~J. and Tao, T. (2007) The {D}antzig selector: {S}tatistical
  estimation when $p$ is much larger than $n$.
\newblock \textit{The Annals of Statistics}, \textbf{35}, 2313--2351.

\bibitem[{Civieta et~al.(2021)Civieta, Aguilera-Morillo and
  Lillo}]{civieta2020adaptive}
Civieta, A.~M., Aguilera-Morillo, M.~C. and Lillo, R.~E. (2021) Adaptive sparse
  group lasso in quantile regression.
\newblock \textit{Advances in Data Analysis and Classification}, \textbf{15},
  547--573.

\bibitem[{Daducci et~al.(2015)Daducci, Dal~Palù, Lemkaddem and
  Thiran}]{daducci2015}
Daducci, A., Dal~Palù, A., Lemkaddem, A. and Thiran, J.~P. (2015) {COMMIT}:
  {Convex} optimization modeling for microstructure informed tractography.
\newblock \textit{IEEE Transactions on Medical Imaging}, \textbf{34}, 246--257.

\bibitem[{Diamond and Boyd(2016)}]{cvxpy-jmlr}
Diamond, S. and Boyd, S. (2016) \pkg{CVXPY}: {A} \proglang{python}-embedded
  modeling language for convex optimization.
\newblock \textit{Journal of Machine Learning Research}, \textbf{17}, 1--5.
\newblock \urlprefix\url{http://jmlr.org/papers/v17/15-408.html}.

\bibitem[{Efron(1986)}]{Efron1986}
Efron, B. (1986) How biased is the apparent error rate of a prediction rule?
\newblock \textit{Journal of the American Statistical Association},
  \textbf{81}, 461--470.

\bibitem[{Friedman et~al.(2022)Friedman, Hastie, Tibshirani, Narasimhan, Tay,
  Simon and Yang}]{R-glmnet}
Friedman, J., Hastie, T., Tibshirani, R., Narasimhan, B., Tay, K., Simon, N.
  and Yang, J. (2022) \textit{\pkg{glmnet}: Lasso and Elastic-Net Regularized
  Generalized Linear Models}.
\newblock \urlprefix\url{https://CRAN.R-project.org/package=glmnet}.
\newblock \proglang{R}~package version~4.1-4.

\bibitem[{Fu et~al.(2020)Fu, Narasimhan and Boyd}]{CVXR}
Fu, A., Narasimhan, B. and Boyd, S. (2020) \pkg{CVXR}: An \proglang{R} package
  for disciplined convex optimization.
\newblock \textit{Journal of Statistical Software}, \textbf{94}, 1--34.

\bibitem[{Fu et~al.(2022)Fu, Narasimhan, Kang, Diamond and Miller}]{R-CVXR}
Fu, A., Narasimhan, B., Kang, D.~W., Diamond, S. and Miller, J. (2022)
  \textit{\pkg{CVXR}: Disciplined Convex Optimization}.
\newblock \urlprefix\url{https://CRAN.R-project.org/package=CVXR}.
\newblock \proglang{R}~package version~1.0-11.

\bibitem[{Gerber et~al.(2017)Gerber, Moesinger and Furrer}]{gerber2017dotcall}
Gerber, F., Moesinger, K. and Furrer, R. (2017) Extending \proglang{R} packages
  to support 64-bit compiled code: An illustration with \pkg{spam64} and
  {GIMMS} {NDVI3g} data.
\newblock \textit{Computer \& Geoscience}, \textbf{104}, 109--119.

\bibitem[{Gerber et~al.(2018)Gerber, Moesinger and Furrer}]{gerber2018dotcall}
--- (2018) \pkg{dotCall64}: An \proglang{R} package providing an efficient
  interface to compiled \proglang{C}, \proglang{C++}, and \proglang{Fortran}
  code supporting long vectors.
\newblock \textit{SoftwareX}, \textbf{7}, 217--221.

\bibitem[{Golub et~al.(1979)Golub, Heath and Wahba}]{golub1979generalized}
Golub, G.~H., Heath, M. and Wahba, G. (1979) Generalized cross-validation as a
  method for choosing a good ridge parameter.
\newblock \textit{Technometrics}, \textbf{21}, 215--223.

\bibitem[{McDonald et~al.(2022)McDonald, Liang, {Solón Heinsfeld} and
  Cohen}]{R-sparsegl}
McDonald, D.~J., Liang, X., {Solón Heinsfeld}, A. and Cohen, A. (2022)
  \textit{\pkg{sparsegl}: Sparse Group Lasso}.
\newblock \urlprefix\url{https://CRAN.R-project.org/package=sparsegl}.
\newblock \proglang{R}~package version~1.0.1.

\bibitem[{Pestilli et~al.(2014)Pestilli, Yeatman, Rokem, Kay and
  Wandell}]{pestilli2014}
Pestilli, F., Yeatman, J.~D., Rokem, A., Kay, K.~N. and Wandell, B.~A. (2014)
  Evaluation and statistical inference for human connectomes.
\newblock \textit{Nature Methods}, \textbf{11}, 1058--1063.

\bibitem[{Qiu and Mei(2022)}]{R-RSpectra}
Qiu, Y. and Mei, J. (2022) \textit{\pkg{RSpectra}: Solvers for Large-Scale
  Eigenvalue and SVD Problems}.
\newblock \urlprefix\url{https://CRAN.R-project.org/package=RSpectra}.
\newblock \proglang{R}~package version~0.16-1.

\bibitem[{{\proglang{R} Core Team}(2022)}]{R-base}
{\proglang{R} Core Team} (2022) \textit{\proglang{R}: A Language and
  Environment for Statistical Computing}.
\newblock \proglang{R} Foundation for Statistical Computing, Vienna, Austria.
\newblock \urlprefix\url{https://www.R-project.org/}.

\bibitem[{Schiavi et~al.(2020)Schiavi, Ocampo-Pineda, Barakovic, Petit,
  Descoteaux, Thiran and Daducci}]{Schiavi2020}
Schiavi, S., Ocampo-Pineda, M., Barakovic, M., Petit, L., Descoteaux, M.,
  Thiran, J.-P. and Daducci, A. (2020) A new method for accurate in vivo
  mapping of human brain connections using microstructural and anatomical
  information.
\newblock \textit{Science Advances}, \textbf{6}, eaba8245.

\bibitem[{Schwarz(1978)}]{Schwarz1978}
Schwarz, G. (1978) Estimating the dimension of a model.
\newblock \textit{The Annals of Statistics}, \textbf{6}, 461--464.

\bibitem[{Simon et~al.(2013)Simon, Friedman, Hastie and
  Tibshirani}]{simon2013sparse}
Simon, N., Friedman, J., Hastie, T. and Tibshirani, R. (2013) A sparse-group
  lasso.
\newblock \textit{Journal of Computational and Graphical Statistics},
  \textbf{22}, 231--245.

\bibitem[{Simon et~al.(2019)Simon, Friedman, Hastie and Tibshirani}]{R-SGL}
--- (2019) \textit{\pkg{SGL}: Fit a GLM (or Cox Model) with a Combination of
  Lasso and Group Lasso Regularization}.
\newblock \urlprefix\url{https://CRAN.R-project.org/package=SGL}.
\newblock \proglang{R}~package version~1.3.

\bibitem[{Tibshirani(1996)}]{tibshirani1996regression}
Tibshirani, R. (1996) Regression shrinkage and selection via the lasso.
\newblock \textit{Journal of the Royal Statistical Society B}, \textbf{58},
  267--288.

\bibitem[{Tibshirani et~al.(2012)Tibshirani, Bien, Friedman, Hastie, Simon,
  Taylor and Tibshirani}]{tibshirani2012strong}
Tibshirani, R., Bien, J., Friedman, J., Hastie, T., Simon, N., Taylor, J. and
  Tibshirani, R.~J. (2012) Strong rules for discarding predictors in lasso-type
  problems.
\newblock \textit{Journal of the Royal Statistical Society B}, \textbf{74},
  245--266.

\bibitem[{Tseng(2001)}]{tseng2001convergence}
Tseng, P. (2001) Convergence of a block coordinate descent method for
  nondifferentiable minimization.
\newblock \textit{Journal of Optimization Theory and Applications},
  \textbf{109}, 475--494.

\bibitem[{Vaiter et~al.(2012)Vaiter, Deledalle, Peyré, Fadili and
  Dossal}]{vaiter2012degrees}
Vaiter, S., Deledalle, C., Peyré, G., Fadili, J. and Dossal, C. (2012) The
  degrees of freedom of the group lasso for a general design.
\newblock \textit{Tech. rep.}, arXiv.

\bibitem[{Van~Essen et~al.(2012)Van~Essen, Ugurbil, Auerbach, Barch, Behrens,
  Bucholz, Chang, Chen, Corbetta, Curtiss, Della~Penna, Feinberg, Glasser,
  Harel, Heath, Larson-Prior, Marcus, Michalareas, Moeller, Oostenveld,
  Petersen, Prior, Schlaggar, Smith, Snyder, Xu and Yacoub}]{vanessen2012}
Van~Essen, D., Ugurbil, K., Auerbach, E., Barch, D., Behrens, T., Bucholz, R.,
  Chang, A., Chen, L., Corbetta, M., Curtiss, S., Della~Penna, S., Feinberg,
  D., Glasser, M., Harel, N., Heath, A., Larson-Prior, L., Marcus, D.,
  Michalareas, G., Moeller, S., Oostenveld, R., Petersen, S., Prior, F.,
  Schlaggar, B., Smith, S., Snyder, A., Xu, J. and Yacoub, E. (2012) The
  {Human} {Connectome} {Project}: {A} data acquisition perspective.
\newblock \textit{NeuroImage}, \textbf{62}, 2222--2231.

\bibitem[{Van~Rossum et~al.(2011)}]{python}
Van~Rossum, G. et~al. (2011) \textit{\proglang{python} Programming Language}.
\newblock \urlprefix\url{https://www.python.org}.

\bibitem[{Wickham(2022)}]{R-testthat}
Wickham, H. (2022) \textit{\pkg{testthat}: Unit Testing for \proglang{R}}.
\newblock \urlprefix\url{https://CRAN.R-project.org/package=testthat}.
\newblock \proglang{R}~package version~3.1.5.

\bibitem[{Wickham et~al.(2022{\natexlab{a}})Wickham, Bryan and
  Barrett}]{R-usethis}
Wickham, H., Bryan, J. and Barrett, M. (2022{\natexlab{a}})
  \textit{\pkg{usethis}: Automate Package and Project Setup}.
\newblock \urlprefix\url{https://CRAN.R-project.org/package=usethis}.
\newblock \proglang{R}~package version~2.1.6.

\bibitem[{Wickham et~al.(2022{\natexlab{b}})Wickham, Chang, Henry, Pedersen,
  Takahashi, Wilke, Woo, Yutani and Dunnington}]{R-ggplot2}
Wickham, H., Chang, W., Henry, L., Pedersen, T.~L., Takahashi, K., Wilke, C.,
  Woo, K., Yutani, H. and Dunnington, D. (2022{\natexlab{b}})
  \textit{\pkg{ggplot2}: Create Elegant Data Visualisations Using the Grammar
  of Graphics}.
\newblock \urlprefix\url{https://CRAN.R-project.org/package=ggplot2}.
\newblock \proglang{R}~package version~3.4.0.

\bibitem[{Wickham et~al.(2022{\natexlab{c}})Wickham, François, Henry and
  Müller}]{R-dplyr}
Wickham, H., François, R., Henry, L. and Müller, K. (2022{\natexlab{c}})
  \textit{\pkg{dplyr}: A Grammar of Data Manipulation}.
\newblock \urlprefix\url{https://CRAN.R-project.org/package=dplyr}.
\newblock \proglang{R}~package version~1.0.10.

\bibitem[{Wickham et~al.(2022{\natexlab{d}})Wickham, Hester, Chang and
  Bryan}]{R-devtools}
Wickham, H., Hester, J., Chang, W. and Bryan, J. (2022{\natexlab{d}})
  \textit{\pkg{devtools}: Tools to Make Developing \proglang{R} Packages
  Easier}.
\newblock \urlprefix\url{https://CRAN.R-project.org/package=devtools}.
\newblock \proglang{R}~package version~2.4.5.

\bibitem[{Wood(2022)}]{R-mgcv}
Wood, S. (2022) \textit{\pkg{mgcv}: Mixed GAM Computation Vehicle with
  Automatic Smoothness Estimation}.
\newblock \urlprefix\url{https://CRAN.R-project.org/package=mgcv}.
\newblock \proglang{R}~package version~1.8-41.

\bibitem[{Wood(2017)}]{woodGams}
Wood, S.~N. (2017) \textit{Generalized Additive Models: {A}n Introduction with
  \proglang{R}}.
\newblock New York: Chapman and Hall/CRC, 2nd edn.

\bibitem[{Xie(2022)}]{R-knitr}
Xie, Y. (2022) \textit{\pkg{knitr}: A General-Purpose Package for Dynamic
  Report Generation in \proglang{R}}.
\newblock \urlprefix\url{https://CRAN.R-project.org/package=knitr}.
\newblock \proglang{R}~package version~1.40.

\bibitem[{Yang and Zou(2015)}]{yang2015fast}
Yang, Y. and Zou, H. (2015) A fast unified algorithm for solving group-lasso
  penalized learning problems.
\newblock \textit{Statistics and Computing}, \textbf{25}, 1129--1141.

\bibitem[{Yang et~al.(2020)Yang, Zou and Bhatnagar}]{R-gglasso}
Yang, Y., Zou, H. and Bhatnagar, S. (2020) \textit{\pkg{gglasso}: Group Lasso
  Penalized Learning Using a Unified BMD Algorithm}.
\newblock \urlprefix\url{https://CRAN.R-project.org/package=gglasso}.
\newblock \proglang{R}~package version~1.5.

\bibitem[{Yuan and Lin(2006)}]{yuan2006model}
Yuan, M. and Lin, Y. (2006) Model selection and estimation in regression with
  grouped variables.
\newblock \textit{Journal of the Royal Statistical Society B}, \textbf{68},
  49--67.

\bibitem[{Zeng and Breheny(2020)}]{zeng2020biglasso}
Zeng, Y. and Breheny, P. (2020) The \pkg{biglasso} package: A memory- and
  computation-effcient solver for lasso model fitting with big data in
  \proglang{R}.
\newblock \textit{The \proglang{R} Journal}, \textbf{12}, 6--19.

\bibitem[{Zeng et~al.(2022)Zeng, Wang and Breheny}]{R-biglasso}
Zeng, Y., Wang, C. and Breheny, P. (2022) \textit{\pkg{biglasso}: Extending
  Lasso Model Fitting to Big Data}.
\newblock \urlprefix\url{https://CRAN.R-project.org/package=biglasso}.
\newblock \proglang{R}~package version~1.5.2.

\bibitem[{Zou et~al.(2007)Zou, Hastie and Tibshirani}]{ZouHastie2007}
Zou, H., Hastie, T. and Tibshirani, R. (2007) {On the `Degrees of Freedom' of
  the Lasso}.
\newblock \textit{The Annals of Statistics}, \textbf{35}, 2173--2192.

\end{thebibliography}

\end{document}